\documentclass[11pt,a4paper]{article}
\usepackage[T1]{fontenc}
\usepackage{lmodern}
\usepackage[a4paper,margin=1in]{geometry}
\usepackage{amsmath,amssymb,amsthm,bm}
\usepackage{graphicx}
\usepackage{booktabs}
\usepackage{placeins}
\usepackage{microtype}
\usepackage[numbers,sort&compress]{natbib}
\usepackage{xcolor}
\usepackage[font=small,labelfont=bf,skip=6pt]{caption}
\usepackage[colorlinks=true,linkcolor=blue!50!black,citecolor=blue!50!black,urlcolor=blue!50!black]{hyperref}

\theoremstyle{plain}
\newtheorem{proposition}{Proposition}

\newcommand{\E}{\mathbb{E}}
\newcommand{\Heff}{\mathbf{H}_\mathrm{eff}}
\newcommand{\tblwidth}{\linewidth}
\newcommand{\tblnote}[2][\scriptsize]{\par\smallskip{#1\raggedright #2\par}}
\newcommand{\Sref}[1]{\ref{#1}}
\newcommand{\Mref}[1]{\ref{#1}}
\hypersetup{pdfauthor={Mohammad Tasin, Muhammad Abdullah Arafat},
            pdftitle={What does a learned post-equalizer correct in AFDM: Pilot-induced Bussgang gain mismatch under power amplifier nonlinearity and phase noise}}

\newcommand{\bDLBA}{1.24}

\newcommand{\bDLTap}{1.25}
\newcommand{\bGBA}{1.66}

\newcommand{\bGL}{4.18}
\newcommand{\bGLR}{2.82}
\newcommand{\bGLRC}{1.90}
\newcommand{\bGMu}{2.53}
\newcommand{\bGTap}{1.90}

\newcommand{\budgetGapNum}{2.0}
\newcommand{\budgetPA}{0.020}
\newcommand{\budgetPN}{0.035}

\newcommand{\causalBer}{0.160}

\newcommand{\cdSixGUncL}{3.6}
\newcommand{\cdSixGUncLR}{5.5}

\newcommand{\cdThreeGTapLR}{$+0.0$}
\newcommand{\cdaSixDLBAL}{0.6}
\newcommand{\cdaSixDLBALR}{1.4}
\newcommand{\cdaSixDLLRL}{0.8}

\newcommand{\cdaSixGBAL}{1.4}

\newcommand{\cdaSixGLRL}{0.1}

\newcommand{\cdaThreeDLBAL}{0.4}
\newcommand{\cdaThreeDLBALR}{1.2}

\newcommand{\cdaThreeGBAL}{3.3}
\newcommand{\cdaThreeGBALR}{1.7}
\newcommand{\cdaThreeGLRL}{1.7}
\newcommand{\cdaThreeGMuL}{2.4}

\newcommand{\cdciSixDLBAL}{$[-1.2,\,-0.2]$}
\newcommand{\cdciSixDLBALR}{$[-2.8,\,-0.7]$}
\newcommand{\cdciSixDLLRL}{$[+0.2,\,+1.9]$}

\newcommand{\cdciSixGBAL}{$[-2.2,\,-0.9]$}

\newcommand{\cdciSixGLRL}{$[-0.6,\,+1.5]$}

\newcommand{\cdciSixGUncL}{$[2.7,\,6.1]$}

\newcommand{\cdciThreeDLBAL}{$[-0.8,\,-0.1]$}
\newcommand{\cdciThreeDLBALR}{$[-2.3,\,-0.7]$}

\newcommand{\cdciThreeGBAL}{$[-3.7,\,-2.9]$}
\newcommand{\cdciThreeGBALR}{$[-2.8,\,-1.2]$}
\newcommand{\cdciThreeGLRL}{$[-2.3,\,-0.3]$}
\newcommand{\cdciThreeGMuL}{$[-2.9,\,-1.9]$}
\newcommand{\cdciThreeGTapLR}{$[-1.3,\,+0.7]$}

\newcommand{\ciIboMax}{21}
\newcommand{\ciIboMaxAll}{26}
\newcommand{\ciRefMax}{4}
\newcommand{\ciRefMaxDD}{18}
\newcommand{\codedCalLR}{10.82}
\newcommand{\codedCalLmmse}{10.68}
\newcommand{\codedIdeal}{7.7}
\newcommand{\codedUncLR}{16.33}
\newcommand{\codedUncLmmse}{14.25}
\newcommand{\epBAMax}{14}
\newcommand{\epBAMin}{4}
\newcommand{\epErrTen}{4\%}
\newcommand{\epErrZero}{12\%}

\newcommand{\epLRMax}{23}
\newcommand{\epLRMin}{7}

\newcommand{\epMuChgMax}{4.8}
\newcommand{\epMuMax}{1.000}
\newcommand{\epMuMin}{0.982}

\newcommand{\epSdTwoZero}{0.014}
\newcommand{\epTapMax}{27}
\newcommand{\epTapMin}{9}

\newcommand{\lrPilThirtyTwo}{2.78}

\newcommand{\muPilThirtyTwo}{2.62}

\newcommand{\nwMax}{2.89}
\newcommand{\nwMin}{2.77}
\newcommand{\offBerEight}{0.000}
\newcommand{\offBerFour}{0.022}
\newcommand{\omegaOne}{24}
\newcommand{\omegaTwo}{14}
\newcommand{\omegaZero}{42}

\newcommand{\pTwoAlg}{2\times10^{-16}}
\newcommand{\pTwoBerAll}{every}
\newcommand{\pTwoBerHi}{6.6}
\newcommand{\pTwoBerLo}{6.5}
\newcommand{\pTwoCrossHi}{6.8}
\newcommand{\pTwoCrossLo}{5.9}
\newcommand{\pTwoJensen}{0.41}
\newcommand{\pTwoLRHi}{6.2}
\newcommand{\pTwoLRLo}{6.0}
\newcommand{\pnRxChgMax}{12}
\newcommand{\poneEightEightg}{0.838}
\newcommand{\poneEightEighto}{0.08}
\newcommand{\poneEightEightp}{0.945}
\newcommand{\poneEightEightv}{0.058}
\newcommand{\poneEightEightw}{0.946}

\newcommand{\poneEightOnew}{0.947}

\newcommand{\poneTwoFiveSixEightw}{0.945}

\newcommand{\poneTwoFiveSixOnew}{0.945}

\newcommand{\rCELR}{14}

\newcommand{\rDLLR}{14}

\newcommand{\rGBA}{60}

\newcommand{\rGLR}{32}
\newcommand{\rGLRC}{54}
\newcommand{\rGMu}{39}
\newcommand{\rGTap}{55}
\newcommand{\randN}{200}
\newcommand{\randWinDL}{2}
\newcommand{\randWinG}{3}
\newcommand{\rcDLBA}{1.24}

\newcommand{\rcDLLR}{2.44}
\newcommand{\rcDLMu}{2.12}
\newcommand{\rcDLRC}{2.79}
\newcommand{\rcGBA}{1.72}

\newcommand{\rcGLR}{2.86}
\newcommand{\rcGMu}{2.56}
\newcommand{\rcGRC}{3.21}
\newcommand{\rcRaw}{0.267}
\newcommand{\rhoCausal}{0.742}
\newcommand{\rhoEnergy}{55}
\newcommand{\rhoFwd}{0.761}

\newcommand{\rhoSym}{0.742}

\newcommand{\robBAminAll}{5}

\newcommand{\tapOverBAmaxSixCE}{12}
\newcommand{\tapOverBAmaxSixDL}{4}
\newcommand{\tapOverBAmaxSixG}{23}
\newcommand{\tdlBAGHi}{71}
\newcommand{\tdlBAGLo}{54}
\newcommand{\tdlFracMax}{10}
\newcommand{\tdlLRDLHi}{17}
\newcommand{\tdlLRDLLo}{12}
\newcommand{\tdlLRGHi}{59}
\newcommand{\tdlLRGLo}{29}
\newcommand{\tdlMuGHi}{64}
\newcommand{\tdlMuGLo}{36}
\newcommand{\tdlPmax}{10^{-5}}
\newcommand{\timeBerEight}{0.43}
\newcommand{\timeBerFortyEight}{0.18}
\newcommand{\timeBerSixty}{0.04}
\newcommand{\timeBerSixtyFour}{0.00}

\newcommand{\winBerForty}{0.064}
\newcommand{\winBerFortyEight}{0.064}

\newcommand{\winBerThirtyTwo}{0.218}
\newcommand{\winBerTwentyFour}{0.253}

\begin{document}

\title{What does a learned post-equalizer correct in AFDM: Pilot-induced Bussgang gain mismatch under power amplifier nonlinearity and phase noise}
\author{\begin{tabular}[t]{@{}c@{}}Mohammad Tasin\thanks{Corresponding author. E-mail: \texttt{tasin2006072@gmail.com}}\end{tabular}%
\hspace{3em}%
\begin{tabular}[t]{@{}c@{}}Muhammad Abdullah Arafat\thanks{E-mail: \texttt{abdullah\_arafat@eee.buet.ac.bd}}\end{tabular}\\[8pt]
Department of Electrical and Electronic Engineering,\\
Bangladesh University of Engineering and Technology (BUET),\\
Dhaka 1000, Bangladesh}
\date{}
\maketitle

\begin{abstract}
\sloppy
Affine frequency division multiplexing (AFDM) is a candidate 6G waveform for fast-moving users. A real power amplifier (PA) distorts strong peaks and the oscillators add phase noise, leaving an error floor after the linear minimum mean-square error (LMMSE) equalizer. An appealing remedy, inspired by reservoir-computing receivers, is a small linear corrector after the equalizer, trained on pilot frames. This paper asks what such a corrector learns. The answer is mostly a single number. If the channel is measured with a separate pilot that is weaker than the data or lacks its peaks, the PA amplifies pilots and data differently, and every equalized symbol comes out shrunk and rotated. The corrector learns to undo this pilot-induced gain mismatch, which can also be computed from a PA model or estimated from received data, without training. With the average PA input 3~dB below saturation and a noise-free low-power pilot, model-based gain and phase correction lowers the uncoded bit error rate by \rGBA\% relative to LMMSE and saves \cdaThreeGBAL~dB at a coded bit error rate of $10^{-3}$, against \rGLR\% and \cdaThreeGLRL~dB for the trained corrector; five 3GPP channel scenarios confirm the uncoded ranking. With a pilot embedded in every frame the mismatch vanishes and the trained corrector only adds errors. The paper also gives an exact condition for when dividing by the true gain lowers the mean-squared error, explains why short-window correctors cannot replace the equalizer, and shows that mis-scaled decoder reliabilities shift coded results by several decibels.
\end{abstract}
\bigskip

% =======================================================
\section{Introduction}
\label{sec:intro}
% =======================================================

\subsection{Background and motivation}

Future 6G networks must serve users moving at hundreds of kilometres per hour, such as high-speed trains, vehicles, drones, and satellites~\cite{afdm6g2025,sui2026mimo}. At such speeds the radio channel spreads every transmitted signal in both time (multipath delay) and frequency (Doppler shift). Orthogonal frequency division multiplexing (OFDM), the waveform of 4G and 5G, then loses the orthogonality of its subcarriers, and its error rate rises. Affine frequency division multiplexing (AFDM)~\cite{bemani2021afdm,bemani2023afdm} addresses this by carrying data on chirps, signals whose frequency sweeps linearly in time. With properly chosen chirp rates, every propagation path lands on a distinct and predictable position after the receiver's discrete affine Fourier transform (DAFT), the chirp counterpart of the Fourier transform used in OFDM. The receiver can therefore separate the paths and combine the energy of all of them (full diversity) while keeping a one-dimensional, OFDM-like transceiver~\cite{bemani2023afdm,afdm6g2025}. Orthogonal time frequency space (OTFS) modulation~\cite{hadani2017otfs,raviteja2018otfs,surabhi2019otfs} offers similar robustness with a two-dimensional structure.

Real transmitters and receivers are not ideal. The power amplifier (PA) compresses large signal peaks, and AFDM, like OFDM, has large peaks. Driving the PA close to saturation, that is, with a small input back-off (IBO), saves power but distorts the signal. The local oscillators add a slowly drifting random phase, known as phase noise. Recent analyses show that multiple-antenna AFDM keeps its full diversity under such hardware impairments when the receiver uses maximum-likelihood detection~\cite{sui2026mimo}. With the low-complexity linear minimum mean-square error (LMMSE) equalizer, however, the impairments still leave an error floor.

Reservoir computing (RC) receivers, recurrent neural networks in which only a linear output layer is trained, learn symbol detection for OFDM and OTFS directly from the received signal, using the few pilot symbols of each subframe~\cite{zhou2020learning,xu2022rcstruct,xu2024twodrc}. A tempting fix in the same spirit is to append a small learned corrector after the equalizer and to train it on a handful of known pilot frames. Such a corrector does reduce the bit error rate (BER) in simulation. This paper asks what the corrector actually learns, because the answer decides whether learning is needed at all.

\subsection{Main findings}

Before it can equalize, a receiver measures the channel with pilot signals. If those pilots have an amplitude distribution different from that of the data, for example a separate sounding frame that carries a single symbol and is too weak to drive the PA into compression, or constant-envelope sequences without peaks, then the PA gives the pilots a different effective gain from the one it gives the data. In effect, the channel is calibrated at a low drive level and used at a higher one. The receiver therefore believes that the channel is stronger than it is for the data, and after equalization every data symbol comes out shrunk (by about 15\% at 3~dB back-off with such a single-symbol sounding) and slightly rotated. For 16-ary quadrature amplitude modulation (16-QAM), a uniform shrinkage moves the outer constellation points three times as far towards the decision thresholds as the inner points, which is costly.

A trained corrector removes most of this shrinkage, and this is nearly all that it does: its weights converge to a single number close to the inverse of the gain error, slightly reduced by the residual error variance (mostly phase noise and PA distortion). The gain error itself follows directly from the Bussgang decomposition of the PA~\cite{bussgang1952,demir2021bussgang}, which writes the PA output as a scaled copy of its input plus uncorrelated distortion, so it can be corrected without any training, either from a PA model or from the received data. When the pilots resemble the data, the mismatch vanishes, and the trained corrector can only add a slight shrinkage and a phase learned from pilot frames that does not match the data frames; it then performs worse than plain LMMSE.

The mismatch is therefore a property of how the channel is measured, not of AFDM or of the PA alone. It is large when the channel is sounded separately from the data at a different drive level, for example with a low-power or constant-envelope preamble, and it is absent when a single pilot is embedded in every data frame, which is the standard AFDM scheme~\cite{bemani2023afdm,yin2022pilot,yin2024diagonally} (Section~\ref{sec:ep}). A further finding concerns coded comparisons. A channel decoder that is fed wrongly scaled bit reliabilities (log-likelihood ratios that ignore the residual distortion) is mismatched, and then a bit-error-rate difference of only 10--21\% between two receivers can appear as a shift of several decibels in the required signal-to-noise ratio.

\subsection{Related work}

\textit{Impaired AFDM.} Sui \textit{et al.}~\cite{sui2026mimo} bound the BER of multiple-input multiple-output (MIMO) AFDM under multiplicative and additive hardware impairments and show that full diversity is retained with maximum-likelihood detection, Huang \textit{et al.}~\cite{huang2026iq} find that receiver in-phase/quadrature (IQ) imbalance degrades AFDM more than OFDM over the additive white Gaussian noise channel, and Gunasekara and Bedeer~\cite{gunasekara2025iq} analyse receiver IQ imbalance and residual carrier frequency offset and compensate them with an extended LMMSE detector. Channel estimation with a single embedded pilot surrounded by guard symbols was introduced for OTFS~\cite{raviteja2019embedded} and developed for AFDM in~\cite{bemani2023afdm,yin2022pilot,yin2024diagonally}. That a memoryless transmitter nonlinearity also distorts the pilots, and hence the channel estimate, has been addressed for single-antenna OFDM by joint channel and nonlinearity estimation~\cite{mokhtari2025joint}, but, to the authors' knowledge, the resulting pilot/data gain mismatch, and what a learned corrector contributes on top of it, have not been quantified for AFDM.

\textit{PA distortion and phase noise.} The Bussgang decomposition~\cite{bussgang1952,demir2021bussgang} is the standard tool for PA distortion in OFDM~\cite{dardari2000,banelli2000}, with the Rapp model~\cite{rapp1991} describing the amplitude compression (AM/AM conversion) of solid-state PAs. Transmit-side digital predistortion based on behavioural PA models, such as the generalized memory polynomial~\cite{morgan2006gmp} or dynamic-deviation-reduction Volterra series~\cite{zhu2006dynamic}, needs PA feedback and is complementary to the receiver-side methods considered here. Common phase error (CPE) estimation is the classical first step against phase noise in OFDM~\cite{wu2004pn,petrovic2007pn}; phase noise has also been analysed for OTFS~\cite{surabhi2019pn}.

\textit{Learned detectors.} Model-driven networks~\cite{shlezinger2020viterbinet,he2020oampnet} and reservoir computing~\cite{zhou2020learning,xu2022rcstruct,xu2024twodrc} learn symbol detection from small amounts of training data, and an echo-state reservoir has also served as a lightweight temporal encoder in a time-series foundation model for key-performance-indicator data of 6G open radio access networks (O-RAN)~\cite{rezazadeh2025reservoir}. The two-dimensional RC (2D-RC) of Xu \textit{et al.}~\cite{xu2024twodrc} exploits the two-dimensional circular structure of the OTFS delay-Doppler channel through circular padding and a sliding two-dimensional window. Whether a windowed readout of this kind can work in the AFDM domain, and what a learned corrector contributes when it follows an LMMSE equalizer, have not been analysed.

\textit{Coded performance.} Bit-interleaved coded modulation is sensitive to the scaling of the log-likelihood ratios (LLRs)~\cite{martinez2009bicm,nguyen2011bicm}. A decoder fed LLRs that ignore residual distortion is mismatched, which matters when receivers with different residual errors are compared.

\subsection{Contributions}

The main contributions of this paper are the following.
\begin{itemize}
\item \textbf{Pilot-induced Bussgang gain mismatch.} The gain mismatch $\mu = B_\mathrm{d}/B_\mathrm{s}$ between the PA gains seen by data and by pilots is identified and quantified for three separate soundings (impulse, constant-envelope, data-like) and for an embedded single pilot. Correcting it alone removes \rGMu\% of the LMMSE bit errors at 3~dB back-off (\rGBA\% together with phase tracking) when the sounding is separate and low-power, and changes the BER by at most \epMuChgMax\% for an embedded pilot (Sections~\ref{sec:system}, \ref{sec:uncoded}, and~\ref{sec:ep}).
\item \textbf{What a trained corrector learns (Proposition~\ref{prop:readout}).} A windowed least-squares readout after LMMSE converges to the scalar $1/(g(1+v))$, the inverse of the post-equalizer gain $g$, shrunk by the factor $1/(1+v)$, where $v$ is the residual error variance; in simulation its other taps carry $\poneEightEighto\%$ of the weight energy.
\item \textbf{An exact correction criterion (Proposition~\ref{prop:threshold}).} Dividing by the oracle gain $g$ lowers the mean-squared error (MSE) if and only if $|1-g|^2 > v(1-|g|^2)$. The resulting crossover back-off depends on the signal-to-noise ratio (SNR) (\pTwoCrossLo--\pTwoCrossHi~dB here); the threshold on the bias is set mainly by phase noise, whereas the bias itself is caused by the PA.
\item \textbf{Why contiguous readouts fail on raw AFDM observations (Proposition~\ref{prop:mismatch}).} For two-path channels, the exact finite-sum structure of the inverse AFDM channel is derived, together with a closed-form mismatch for any readout restricted to a set of diagonals. Contiguous readouts invert the reference channel only if they span nearly the whole frame (a BER of \winBerFortyEight{} with 48 of 64 DAFT-domain taps and zero with 56; \timeBerSixty{} with 60 time-domain samples), whereas eight taps placed on the channel-dependent diagonals give no observed errors.
\item \textbf{Training-free receivers and fragile coded comparisons.} Gain correction with decision-directed CPE tracking beats the trained corrector, with or without CPE tracking, in every simulated row of the robustness study with a correct PA model, and a blind one-tap gain tracker beats the trained corrector without tracking in every row; both also beat it on five tapped-delay-line (TDL) channel configurations of the 3rd Generation Partnership Project (3GPP). Against the trained corrector with tracking, the blind tracker is competitive but not uniformly better. With calibrated LLRs the trained corrector shows no detectable coded gain at 6~dB back-off, and uncalibrated LLRs shift coded results by several decibels.
\end{itemize}

\subsection{Organization and notation}

Section~\ref{sec:system} introduces the system model and the gain mismatch, Section~\ref{sec:receivers} describes the receivers, Section~\ref{sec:theory} analyses what a trained readout learns, when gain correction helps, and why short-window readouts fail on raw observations, Section~\ref{sec:results} presents the numerical results, Section~\ref{sec:discussion} discusses design guidelines and limitations, and Section~\ref{sec:conclusion} concludes. The Supplementary Material contains the complete simulation set-up, full proofs and validations, and additional results; its sections, tables and equations are numbered with the prefix S.

Bold lowercase (uppercase) letters denote vectors (matrices), except the DAFT-domain symbol vector $\mathbf{X}$; $(\cdot)^*$ and $(\cdot)^H$ denote conjugate and conjugate transpose; $\mathbb{E}[\cdot]$ is the expectation; $(a)_N$ is $a$ modulo $N$; $\delta_{ij}$ is the Kronecker delta; and $\mathcal{CN}(0,\sigma^2)$ is the circularly symmetric complex Gaussian distribution.

% =======================================================
\section{System model and the gain mismatch}
\label{sec:system}
% =======================================================

\subsection{AFDM, channel, and impairments}

An AFDM frame carries $N$ symbols $\mathbf{X}\in\mathcal{A}^N$ from a unit-energy Gray-coded 16-QAM alphabet $\mathcal{A}$. The DAFT is $\mathbf{A}=\mathbf{\Lambda}_{c_2}\mathbf{F}\mathbf{\Lambda}_{c_1}$, where $\mathbf{F}$ is the unitary discrete Fourier transform (DFT) matrix and $\mathbf{\Lambda}_c=\mathrm{diag}(e^{-j2\pi c n^2})_{n=0}^{N-1}$~\cite{bemani2023afdm}. The transmitter forms $\mathbf{x}=\mathbf{A}^H\mathbf{X}$ and prepends a chirp-periodic prefix (CPP) of length $L_\mathrm{CPP}$. The simulations use $N=64$, $L_\mathrm{CPP}=16$, $c_1=(2(\nu_\mathrm{max}+1)+1)/(2N)=9/128$ with $\nu_\mathrm{max}=3$, $c_2=1/128$, and a sampling rate $f_s=1.92$~MHz (30~kHz spacing).

\textit{Channel.} A $P$-path channel with integer delays $l_i$, Doppler shifts $\nu_i$ (in subcarrier spacings), and gains $h_i$ acts on the transmitted samples $s[n]$ as
\begin{equation}
r[n]=\sum_{i=1}^{P} h_i\, e^{j2\pi\nu_i n/N}s[n-l_i]+w[n],
\label{eq:channel}
\end{equation}
with $w[n]\sim\mathcal{CN}(0,N_0)$. After CPP removal and the DAFT, path $i$ maps symbol $X[(m+\delta_i)_N]$ onto output $m$, with
\begin{equation}
\delta_i=(2Nc_1l_i-\nu_i)_N ,
\label{eq:offset}
\end{equation}
so the effective channel $\Heff=\sum_i h_i\mathbf{A}_i$ is a sum of unitary \emph{monomial} matrices $\mathbf{A}_i$ (one non-zero entry per row and column, on the $\delta_i$-th cyclic diagonal)~\cite{bemani2023afdm}. The sign of $\nu_i$ in~\eqref{eq:offset} follows from the $e^{+j2\pi\nu n/N}$ convention above; it is opposite in~\cite{bemani2023afdm}. The reference channel used throughout has $h=[1,\,0.7+0.3j]$, $l=[0,3]$, and $\nu=[0,2]$, giving $\delta_2=27-2=25$. With 30~kHz spacing, $\nu_2=2$ corresponds to a 60~kHz Doppler shift, far above terrestrial vehicle speeds; it stresses the diagonal structure, whereas the 3GPP TDL channels of Section~\ref{sec:robust} use realistic speeds. Section~\ref{sec:results} also adds fractional Doppler and random geometries.

\textit{PA.} The memoryless Rapp amplitude model~\cite{rapp1991}, extended by a Saleh-type amplitude-to-phase (AM/PM) conversion term~\cite{saleh1981},
\begin{equation}
\begin{aligned}
f(x)&=x\,G_\mathrm{R}(|x|)\,e^{j\varphi(|x|)},\\
G_\mathrm{R}(r)&=\big(1+(r/A_\mathrm{sat})^{2p}\big)^{-1/(2p)},\\
\varphi(r)&=\alpha_\mathrm{PM}\,r^2/(A_\mathrm{sat}^2+r^2),
\end{aligned}
\label{eq:rapp}
\end{equation}
is applied to the time samples, with $p=2$ and $\alpha_\mathrm{PM}=0.1$~rad; the IBO is $10\log_{10}(A_\mathrm{sat}^2/\mathbb{E}|x|^2)$. The PA input is scaled by the back-off and the output is rescaled by its inverse, so the small-signal gain is one and the SNR is referenced to the nominal linear output power. This isolates the distortion from the efficiency cost of back-off, which is not modelled.

\textit{Phase noise.} The transmit oscillator adds a Wiener phase $\theta[n]=\theta[n-1]+\Delta[n]$, $\Delta[n]\sim\mathcal{N}(0,2\pi\Delta f/f_s)$, after the PA. With a linewidth of $\Delta f=100$~Hz the phase standard deviation accumulated over one 64-sample block is $8.3^\circ$. Placing the oscillator at the receiver changes the BER of every receiver except the decision-directed distortion-cancellation (DD-DC) reference of Section~\ref{sec:receivers} by at most \pnRxChgMax\% (impulse sounding; Section~\ref{sec:robust}).

\subsection{Bussgang model and post-equalizer gain}

For Gaussian-like AFDM samples the PA output decomposes as $f(x)=B_\mathrm{d}x+d$ with $B_\mathrm{d}=\mathbb{E}[f(x)x^*]/\mathbb{E}|x|^2$ and $\mathbb{E}[dx^*]=0$; $\sigma_d^2=\mathbb{E}|d|^2$ is the distortion variance~\cite{bussgang1952,demir2021bussgang}. Let $\mathbf{H}$ be the time-domain channel after CPP removal, $\Heff=\mathbf{A}\mathbf{H}\mathbf{A}^H$ the effective channel without phase noise, and $\mathbf{\Phi}=\mathrm{diag}(e^{j\theta[n]})$. The DAFT-domain received frame is
\begin{equation}
\begin{split}
\mathbf{y}={}&B_\mathrm{d}\Heff\mathbf{X}+B_\mathrm{d}\mathbf{A}\mathbf{H}(\mathbf{\Phi}-\mathbf{I})\mathbf{A}^H\mathbf{X}\\
&+\mathbf{A}\mathbf{H}\mathbf{\Phi}\mathbf{d}+\tilde{\mathbf{w}},
\end{split}
\label{eq:rx}
\end{equation}
where $\tilde{\mathbf{w}}$ is white noise. The first term is the useful signal with the data gain $B_\mathrm{d}$, the second is phase-noise interference (a common phase error plus inter-carrier interference), and the third is filtered PA distortion. The phase process is applied to the CPP-extended samples, so the small loss of circularity it causes is part of the simulated interference.

\textit{Channel estimation and the gain mismatch.} The receiver estimates the channel from pilots that pass through the same PA. A pilot waveform with a different amplitude distribution sees a different gain $B_\mathrm{s}$, so the estimate is $\hat{\mathbf{H}}_\mathrm{eff}\approx B_\mathrm{s}\Heff$, and the LMMSE equalizer
\begin{equation}
\mathbf{W}=(\hat{\mathbf{H}}_\mathrm{eff}^H\hat{\mathbf{H}}_\mathrm{eff}+N_0\mathbf{I})^{-1}\hat{\mathbf{H}}_\mathrm{eff}^H
\label{eq:lmmse}
\end{equation}
returns $\mathbf{z}=\mathbf{W}\mathbf{y}$ with an average symbol gain close to
\begin{equation}
\mu=\frac{B_\mathrm{d}}{B_\mathrm{s}} .
\label{eq:mu}
\end{equation}
Three separate pilot (sounding) types are considered. \emph{Impulse sounding (G)}: one DAFT impulse per frame, whose time samples have the tiny constant amplitude $1/\sqrt{N}$, so the PA is linear and $B_\mathrm{s}\approx 1$; the sounding is noise-free, but each frame sees its own phase-noise path. It is close to the perfect (``genie'') channel knowledge assumed in much of the literature. \emph{Constant-envelope pilots (CE)}: the same impulses scaled to the data power, so every sample has $|s|=1$; $B_\mathrm{s}$ is the PA gain at that fixed amplitude, and 256 noisy frames are averaged. \emph{Data-like pilots (DL)}: 256 random 16-QAM frames pass through the PA and the channel is obtained by least squares, so $B_\mathrm{s}=B_\mathrm{d}$. The practical alternative, a single \emph{embedded pilot} in every data frame, is treated in Section~\ref{sec:ep}.

\begin{table}[htbp]\centering\small
\caption{Data Bussgang gain and pilot-induced gain mismatch $\mu$ versus input back-off.}
\label{tab:mu}
\begin{tabular*}{\tblwidth}{@{\extracolsep{\fill}}lcccccccc@{}}
\toprule
IBO (dB) & 1 & 2 & 3 & 4 & 5 & 6 & 7 & 8\\
\midrule
$|B_\mathrm{d}|$ & 0.765 & 0.809 & 0.850 & 0.885 & 0.914 & 0.938 & 0.956 & 0.970\\
$\sigma_d^2/|B_\mathrm{d}|^2$ & 0.038 & 0.028 & 0.019 & 0.013 & 0.008 & 0.005 & 0.003 & 0.002\\
$|\mu|$, G & 0.765 & 0.809 & 0.850 & 0.885 & 0.914 & 0.938 & 0.956 & 0.970\\
$\angle\mu$, G ($^\circ$) & 2.9 & 2.6 & 2.4 & 2.1 & 1.9 & 1.7 & 1.4 & 1.2\\
$|\mu|$, CE & 0.864 & 0.880 & 0.899 & 0.918 & 0.936 & 0.952 & 0.966 & 0.976\\
$|\mu|$, DL & 1 & 1 & 1 & 1 & 1 & 1 & 1 & 1\\
\bottomrule
\end{tabular*}
\tblnote[\footnotesize]{Rapp PA with $p=2$ and $\alpha_\mathrm{PM}=0.1$~rad. G: impulse sounding; CE: constant-envelope pilots; DL: data-like pilots.}
\end{table}

\begin{figure}[t]
  \centering
  \includegraphics[width=\linewidth]{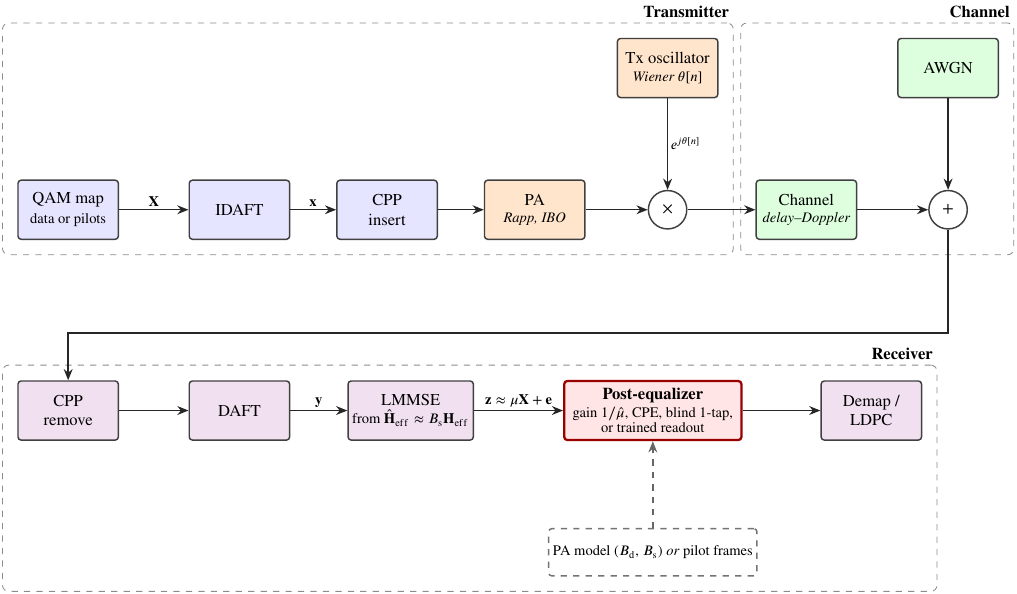}
  \caption{AFDM link with PA nonlinearity, phase noise, and the post-equalizer.}
  \label{fig:blockdiag}
\end{figure}

Table~\ref{tab:mu} lists the mismatch. With impulse sounding the data symbols are shrunk by 15\% at 3~dB back-off, a systematic error comparable to the distortion power itself (an error power of $0.15^2=0.023$ against $\sigma_d^2/|B_\mathrm{d}|^2=0.019$). Constant-envelope pilots reduce the error by about a third at 3~dB back-off (from 15\% to 10\%) but do not remove it; data-like pilots remove it by construction. Fig.~\ref{fig:blockdiag} summarizes the link and the location of the post-equalizer.

% =======================================================
\section{Receivers under test}
\label{sec:receivers}
% =======================================================

All receivers start from the pilot-based channel estimate $\hat{\mathbf{H}}_\mathrm{eff}$. LMMSE/$\hat\mu$, the blind one-tap tracker, and LR post-process the LMMSE output $\mathbf{z}$ of~\eqref{eq:lmmse}, whereas BA+CPE and DD-DC use their own filters.

\textit{Gain correction (LMMSE/$\hat\mu$).} The output is $\mathbf{z}/\hat\mu$, with $\hat\mu$ from~\eqref{eq:mu} evaluated on the nominal PA model (for G and CE pilots, $B_\mathrm{s}$ follows from the known pilot amplitude; $B_\mathrm{d}$ is a one-dimensional integral or a look-up table).

\textit{Bussgang-aware LMMSE with CPE tracking (BA+CPE).} The channel $\mathbf{H}_0=\hat{\mathbf{H}}_\mathrm{eff}/B_\mathrm{s}$ is rescaled by $B_\mathrm{d}$, and the distortion is treated as coloured noise:
\begin{equation}
\mathbf{W}_\mathrm{BA}=B_\mathrm{d}^*\mathbf{H}_0^H\big((|B_\mathrm{d}|^2+\sigma_d^2)\mathbf{H}_0\mathbf{H}_0^H+N_0\mathbf{I}\big)^{-1},
\label{eq:wba}
\end{equation}
normalized to unit diagonal gain. A decision-directed variant of standard CPE correction~\cite{wu2004pn,petrovic2007pn}, $\hat\phi=\angle\sum_m z[m]\check X^*[m]$ (two iterations per frame), where $\check X[m]$ is the hard 16-QAM decision on $z[m]$, then removes the per-frame CPE.

\textit{Blind one-tap.} No PA model is used: per frame, the gain $\hat g=\sum_m z[m]\check X^*[m]/\sum_m|\check X[m]|^2$ is estimated from the hard decisions $\check X[m]$, and the output is $\mathbf{z}/\hat g$ (two iterations).

\textit{Trained linear readout (LR).} A shared $N_w$-tap filter over the cyclic window $[z[m-N_w+1],\dots,z[m]]$ is fitted by ridge least squares to 8 known pilot frames ($N_w=8$, regularization $10^{-4}\,\mathrm{tr}(\mathbf{U}^H\mathbf{U})/N_w$, where $\mathbf{U}$ stacks the windows of all pilot symbols). It represents linear post-LMMSE correctors, including the readout of a reservoir driven by the equalizer output when that reservoir operates in its linear regime. \emph{LR+CPE} follows it with the same decision-directed phase estimate as BA+CPE, which gives the learned corrector the per-frame tracking it otherwise lacks.

\textit{Decision-directed distortion cancellation (DD-DC, reference).} Starting from BA+CPE decisions, the PA distortion $f(\hat{\mathbf{x}})-B_\mathrm{d}\hat{\mathbf{x}}$ is regenerated with the \emph{exact} PA curve, passed through $\mathbf{H}_0$, and subtracted, and the frame is re-equalized (two iterations). It requires full PA knowledge and serves as an upper reference for the uncoded BER.

Table~\ref{tab:complexity} compares the per-frame cost. LMMSE$/\hat\mu$, the blind one-tap tracker, and BA+CPE add at most about $0.6N^2$ multiplications to LMMSE, DD-DC costs several times more, and LR additionally needs pilot frames and a training step.

\begin{table}[htbp]\centering\small
\caption{Per-frame complexity and side information.}
\label{tab:complexity}
\footnotesize
\setlength{\tabcolsep}{2pt}
\begin{tabular*}{\tblwidth}{@{\extracolsep{\fill}}llcc@{}}
\toprule
Receiver & Side information & Training & Mult./frame\\
\midrule
LMMSE & $\hat{\mathbf{H}}_\mathrm{eff}$ & -- & $N^2=4096$\\
LMMSE/$\hat\mu$ & PA model (scalar) & -- & $\approx4.2$k\\
BA+CPE & PA model ($B_\mathrm{d},\sigma_d^2$) & -- & $\approx6.5$k\\
Blind one-tap & none & -- & $\approx6.4$k\\
Linear readout (LR) & 8 pilot frames & ridge LS & $\approx4.6$k$^\dagger$\\
DD-DC (ref.) & exact PA curve & -- & $\approx27$k\\
\bottomrule
\end{tabular*}
\tblnote{Mult./frame: complex multiplications per frame for $N=64$ and $|\mathcal{A}|=16$. $^\dagger$Plus $\approx3.3\times10^4$ per training (once per coherence interval).}
\end{table}

% =======================================================
\section{Analysis}
\label{sec:theory}
% =======================================================

\subsection{What a trained readout learns}

The post-LMMSE output per symbol is written as
\begin{equation}
z[m]=g\,X[m]+e[m],\qquad \mathbb{E}\big[e[m]X^*[m]\big]=0,
\label{eq:zmodel}
\end{equation}
where $g=\mathbb{E}[zX^*]$ is the effective gain and $e$ collects distortion, phase-noise interference, and noise. The gain satisfies $g\approx\mu$; at 3~dB back-off with impulse sounding $|g|=\poneEightEightg$ against $|\mu|=0.850$, the difference being LMMSE noise shrinkage and phase-noise loss. The orthogonality in~\eqref{eq:zmodel} holds by the definition of $g$, for any input distribution. Let $v=\mathbb{E}|e|^2/|g|^2$ denote the residual variance after bias removal.

\begin{proposition}[Trained readout]
\label{prop:readout}
Let $X$ be i.i.d.\ and zero-mean with $\mathbb{E}|X|^2=1$, let the gain $g$ be the same for all $m$, let $\mathbb{E}[e[m]X^*[m']]=0$ for all $m,m'$, and let $e$ be zero-mean and white, $\mathbb{E}[e[m]e^*[m']]=|g|^2v\,\delta_{mm'}$. As the number of pilot symbols grows, the least-squares readout with any window $N_w$ converges to the single tap
\begin{equation}
w_\infty=\frac{1}{g\,(1+v)},
\label{eq:winf}
\end{equation}
all other taps vanish, and its output $w_\infty z[m]=X[m]/(1+v)+e[m]/(g(1+v))$ has MSE $v/(1+v)$ and a residual scale factor of $1/(1+v)$. The bias-free correction $z/g$ has MSE $v$ and no scale error.
\end{proposition}

\begin{proof}
The normal equations of the windowed least-squares problem involve $\mathbb{E}[z[m-j]X^*[m]]=g\,\delta_{j0}$ and $\mathbb{E}[z[m-j]z^*[m-i]]=|g|^2(1+v)\delta_{ij}$ under the stated assumptions, so the solution is $w_j=\delta_{j0}/(g(1+v))$. The MSE follows by substitution (Supplementary Section~\Sref{sec:p1proof}).
\end{proof}

After LMMSE the gain varies slightly with $m$ and the PA distortion is slightly coloured, so the assumptions hold only approximately; the measured effect is small. At 3~dB back-off and 24~dB $E_b/N_0$ the measured residual variance is $v=\poneEightEightv$ (100 runs), so~\eqref{eq:winf} predicts $|w_\infty g|=\poneEightEightp$. The fitted LR gives $\poneEightEightw$ with 8 pilot frames and $\poneTwoFiveSixEightw$ with 256 frames, and with 8 frames the seven other taps carry $\poneEightEighto\%$ of the weight energy. A trained readout therefore learns a slightly shrunk inverse gain. The shrinkage minimizes the MSE but, for QAM with fixed decision thresholds, scales every symbol by $1/(1+v)$, about 5\% too small, which is one reason why LR loses to the unbiased correction in BER even with 32 pilot frames (\lrPilThirtyTwo{} against \muPilThirtyTwo{} $\times10^{-2}$; Section~\ref{sec:robust}). With finite pilots it also inherits the frame-to-frame CPE of the pilot frames, which do not match those of the data frames. If the pilots resemble the data ($\mu=1$), there is nothing to learn, and LR can only add these two penalties.

\subsection{When does gain correction help?}

\begin{proposition}[Correction criterion]
\label{prop:threshold}
For the model~\eqref{eq:zmodel} with $\mathbb{E}|X|^2=1$,
\begin{equation}
\mathrm{MSE}(z)-\mathrm{MSE}(z/g)=|1-g|^2-v\,(1-|g|^2).
\label{eq:identity}
\end{equation}
Hence bias removal lowers the MSE if and only if $|1-g|^2>v(1-|g|^2)$. For real $g=1-\eta$ with $\eta>0$ this is $\eta>2v/(1+v)\approx2v$; for $\eta<0$ ($g>1$) bias removal always lowers the MSE.
\end{proposition}

\begin{proof}
By orthogonality, $\mathrm{MSE}(z)=\mathbb{E}|(g-1)X+e|^2=|1-g|^2+|g|^2v$ and $\mathrm{MSE}(z/g)=\mathbb{E}|e/g|^2=v$. Subtracting gives~\eqref{eq:identity}. For $g=1-\eta$, $\eta^2>v(2\eta-\eta^2)$ is equivalent to $\eta(1+v)>2v$ for $\eta>0$.
\end{proof}

\begin{figure}[t]
\centering
\includegraphics[width=\linewidth]{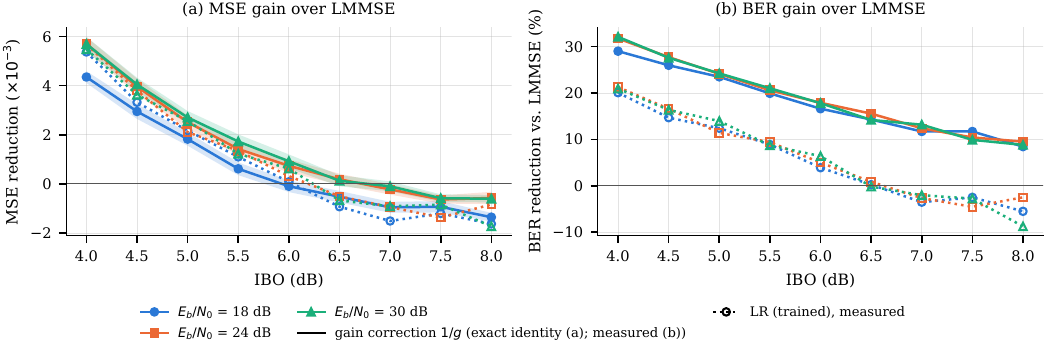}
\caption{Validation of Proposition~\ref{prop:threshold} with impulse sounding: (a) MSE and (b) BER reduction relative to LMMSE.}
\label{fig:threshold}
\end{figure}

The identity is algebraic, so it contains no fitted constant. Here $g$ is the data-aided least-squares gain, an oracle quantity that a receiver would have to estimate; estimating it adds a penalty that is quantified in Supplementary Section~\Sref{sec:p2proof}. The first-order form $\eta>2v$ shows the trade-off clearly: dividing by $g$ removes the bias $\eta^2$ but amplifies the residual from $|g|^2v$ to $v$. Fig.~\ref{fig:threshold}(a) evaluates~\eqref{eq:identity} run by run on 100 independent runs per point and averages (solid lines; the shaded bands are 95\% intervals of the mean). The MSE gain of the correction changes sign between \pTwoCrossLo{} and \pTwoCrossHi~dB back-off, depending on the SNR, so the crossover is an SNR-dependent range rather than a single back-off value. The measured MSE gain of LR crosses zero in nearly the same range (\pTwoLRLo--\pTwoLRHi~dB). Proposition~\ref{prop:readout} implies that a readout trained on unlimited, phase-matched pilots could never lose to LMMSE in MSE, since $v/(1+v)$ is below both $\mathrm{MSE}(z)$ and $v$. The measured crossover of LR therefore reflects the finite pilot set and the mismatch between the phase of the pilot frames and that of each data frame, and Proposition~\ref{prop:threshold} applies to the oracle correction $z/g$ rather than to the trained readout. Evaluating~\eqref{eq:identity} once at the run-averaged $g$ and $v$ differs from the run average by at most $\pTwoJensen\times10^{-3}$, because $g$ and $v$ vary from run to run, whereas the per-run identity holds to $\pTwoAlg$.

Two further facts matter for the design. First, the residual variance decomposes additively (to within \budgetGapNum\%) into noise, phase-noise, and PA terms; at 24~dB the phase-noise part ($\budgetPN$) does not depend on the back-off and exceeds the PA part ($\budgetPA$ at 3~dB back-off, less above it), so phase noise sets the crossover more than the PA does. Second, Fig.~\ref{fig:threshold}(b) shows that, in BER, bias removal helped at \pTwoBerAll{} back-off and SNR tested (IBO 4--8~dB, $E_b/N_0$ of 18, 24, and 30~dB; impulse sounding, 16-QAM), even where~\eqref{eq:identity} is negative, because a uniform scale error moves all 16-QAM outer points towards the decision boundaries. The trained LR, which removes the bias only partially and adds estimation error, crosses zero BER gain at \pTwoBerLo--\pTwoBerHi~dB.

\subsection{Why contiguous readouts fail on raw AFDM observations}

Reservoir computing detectors for OTFS~\cite{xu2024twodrc} apply a readout shared across symbol positions to a local window of received samples. To the authors' knowledge no such detector has been proposed for AFDM; the adaptations tested below were constructed for this study and are not published baselines. On raw AFDM observations a windowed readout must approximate $\Heff^{-1}$. For two paths, $\Heff=h_1\mathbf{A}_1(\mathbf{I}+c\mathbf{M})$ with $c=h_2/h_1$ and $\mathbf{M}=\mathbf{A}_1^{-1}\mathbf{A}_2$, a unitary monomial matrix on cyclic diagonal $\Delta s=(\delta_2-\delta_1)_N$.

\begin{proposition}[Inverse structure]
\label{prop:mismatch}
If $|c|<1$, then $\mathbf{M}^N$ is diagonal with unit-modulus diagonal entries and, exactly,
\begin{equation}
\begin{split}
\Heff^{-1}={}&\frac{1}{h_1}\Big[\sum_{k=0}^{N-1}(-c)^k\mathbf{M}^k\Big]\\
&\times\big(\mathbf{I}-(-c)^N\mathbf{M}^N\big)^{-1}\mathbf{A}_1^{-1}.
\end{split}
\label{eq:neumann}
\end{equation}
The $k$-th term occupies cyclic diagonal $(k\Delta s-\delta_1)_N$ (shifted by $-\delta_1$, the diagonal of $\mathbf{A}_1^{-1}$) and carries the energy fraction $\omega_k=(1-|c|^2)|c|^{2k}/(1-|c|^{2N})$, exactly so when $\gcd(\Delta s,N)=1$ (the $N$ terms then occupy distinct diagonals). For a readout restricted to a set $\mathcal{B}$ of diagonals, the relative Frobenius error of the best approximation of $\Heff^{-1}$ is
\begin{equation}
\rho_\mathcal{B}=\Big(1-\sum_{k:\,(k\Delta s-\delta_1)_N\in\mathcal{B}}\omega_k\Big)^{1/2}.
\label{eq:rho}
\end{equation}
\end{proposition}

\begin{proof}
$\mathbf{M}$ maps each index to the next index of a cyclic orbit with step $\Delta s$, so $\mathbf{M}^N$ returns every index to itself and is diagonal. Multiplying $(\mathbf{I}+c\mathbf{M})$ by $\sum_{k=0}^{N-1}(-c)^k\mathbf{M}^k$ telescopes to $\mathbf{I}-(-c)^N\mathbf{M}^N$, which is invertible for $|c|<1$. Each $\mathbf{M}^k$ is unitary and monomial on diagonal $(k\Delta s)_N$, and for $\gcd(\Delta s,N)=1$ the factor $(\mathbf{I}-(-c)^N\mathbf{M}^N)^{-1}$ is a scalar multiple of the identity (all diagonal entries of $\mathbf{M}^N$ are equal, being products around the single orbit), which gives the energy fractions; for $\gcd(\Delta s,N)>1$ the terms on the same diagonal are summed (Supplementary Section~\Sref{sec:p3proof}). The best Frobenius approximation within a set of diagonals keeps those diagonals and discards the rest, which gives~\eqref{eq:rho}.
\end{proof}

\begin{figure}[t]
\centering
\includegraphics[width=\linewidth]{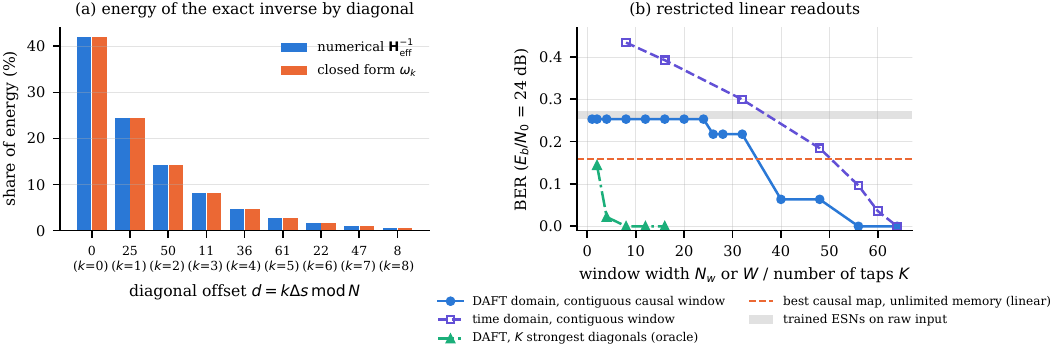}
\caption{Structure of $\Heff^{-1}$ (Proposition~\ref{prop:mismatch}): (a) energy per cyclic diagonal and (b) BER of restricted readouts.}
\label{fig:prop3}
\end{figure}

For the reference channel ($|c|=0.762$, $\Delta s=25$, $\gcd(25,64)=1$) every one of the 64 diagonals is occupied. Fig.~\ref{fig:prop3}(a) shows that the closed form matches the numerical inverse: \omegaZero\% of the energy lies on the main diagonal, and \omegaOne\% and \omegaTwo\% on diagonals 25 and 50. For a causal 8-lag window (diagonals $0,-1,\dots,-7$) only $k=0$ and a few terms with $k\geq5$ fall inside the band, so $\rho_\mathcal{B}=\rhoCausal$, that is, \rhoEnergy\% of the \emph{energy} of the inverse lies outside the window; a symmetric band of $\pm7$ diagonals (15 taps) gives the same value, $\rhoSym$, and a forward-only 8-lag window gives $\rhoFwd$. Eq.~\eqref{eq:rho} matches numerical projections to $10^{-13}$ over 168 two-path geometries (symmetric $\pm7$ band). Its value lies between about $0.8|c|$ and $|c|$, where $|c|$ is the weak-to-strong path ratio, whenever $\Delta s$ falls outside the band, and it drops sharply when $\Delta s$ falls inside it (39 of the 168 geometries), so the mismatch depends on the channel geometry (Supplementary Table~\Sref{tab:geo}).

What does this mean for detection? Fig.~\ref{fig:prop3}(b) shows the BER of the best \emph{linear} equalizers with restricted support, all computed with noise only (no hardware impairments). With a causal window of width $N_w$, the BER stays at \winBerTwentyFour{} up to $N_w=24$, falls to \winBerThirtyTwo{} at 32 and \winBerForty{} at 40--48, and is zero at $N_w=56$ of $N=64$ (Supplementary Table~\Sref{tab:causal}). Even an unlimited-memory causal map, the most that a forward-running reservoir can implement in its linear regime, reaches only \causalBer. These values are benchmarks for linear maps and do not bound nonlinear reservoirs; the empirical evidence for those is that trained echo-state networks (ESNs) on raw observations reach 0.25--0.27 for reservoir sizes 32--512, spectral radii 0.9--0.99, and 8--256 pilot frames (grey band in Fig.~\ref{fig:prop3}(b)), and the 2D-RC of~\cite{xu2024twodrc} adapted to AFDM reaches \rcRaw{} (both at 3~dB back-off with the default PA and phase noise). The failure is thus structural for contiguous windows, which is why a linear readout needs an equalizer that inverts $\Heff$ in full first.

Two limits of this result should be stated. First, it concerns \emph{contiguous} windows. A readout whose $K$ taps sit on the $K$ strongest diagonals of the inverse has $\rho_\mathcal{B}\approx|c|^K$ and reaches a BER of \offBerFour{} at $K=4$ and \offBerEight{} at $K=8$ (Fig.~\ref{fig:prop3}(b)); but these diagonals depend on the channel through $\Delta s=(2Nc_1\Delta l-\Delta\nu)_N$ and must be known or estimated, so such a readout is a model-based sparse inverse, not a generic reservoir. Second, moving to the time domain does not help: the inverse DAFT spreads every symbol over all $N$ samples with equal magnitude, and the best linear estimator restricted to $W$ consecutive time samples has a BER of \timeBerEight{} at $W=8$ and \timeBerFortyEight{} at $W=48$, and approaches zero only for $W$ close to $N$ (\timeBerSixty{} at $W=60$, \timeBerSixtyFour{} at $W=64$; Fig.~\ref{fig:prop3}(b)).

% =======================================================
\section{Numerical results}
\label{sec:results}
% =======================================================

\subsection{Simulation set-up}

Unless stated otherwise, the results use the reference two-path channel, 3~dB back-off, 24~dB $E_b/N_0$, 8 pilot frames followed by 64 data frames, and 100 independent runs per point (new data, noise, and phase noise in every run). Every table row and figure point uses the same seeds and identical frames for all receivers, so all comparisons are paired. $E_b/N_0$ is defined per information bit, $N_0=1/(\log_2|\mathcal{A}|\,R\,10^{E_b/N_0/10})$ with $R=1$ uncoded, and is referenced to the nominal linear PA output. Uncoded uncertainty is reported as 95\% bootstrap intervals over runs: in the reference configuration the half-width is at most \ciRefMax\% of the mean for every receiver except the DD-DC reference (\ciRefMaxDD\%), and at most \ciIboMax\% anywhere in Fig.~\ref{fig:ibo} for the other receivers (\ciIboMaxAll\% including DD-DC); the largest values occur where the BER is below $10^{-3}$.

Coded results use the 5G new radio (NR) low-density parity-check (LDPC) code~\cite{richardson2018ldpc,3gpp38212} (base graph~2, lifting size 44, $k=324$, $n=648$, rate 1/2, 20 belief-propagation iterations) with three AFDM frames per codeword and 10 runs of 100 codewords per point; intervals are bootstrapped over the 10 runs. Each 648-bit codeword is padded with 120 random bits to fill the three frames; this overhead, about 0.7~dB, is not charged to $E_b/N_0$ and is the same for all receivers. Unless noted otherwise, LLRs are max-log values scaled by the residual variance $\mathbb{E}|\hat X-X|^2$ that each receiver measures on its own pilot frames.

\textit{Software and tools.} The AFDM link is simulated with a purpose-built NumPy implementation that is shared by all experiments; the 5G NR LDPC codec and the 3GPP TDL channel models are taken from Sionna 1.2.2~\cite{hoydis2022sionna}. Parts of the simulation and analysis code, the scripts that generate the figures and most tables and in-text numbers from the stored per-run results, and automated cross-checks of the reported numbers were developed with the assistance of the large language model Claude (Anthropic; Sonnet and Opus model families, 2026). All results were checked by the authors. The complete parameter list is given in Supplementary Table~\Sref{tab:params}.

\subsection{Uncoded performance and the role of the pilots}
\label{sec:uncoded}

\begin{figure}[t]
\centering
\includegraphics[width=\linewidth]{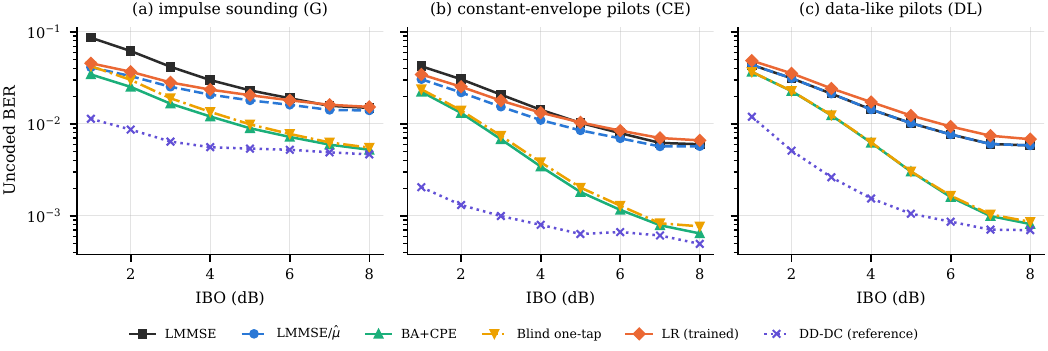}
\caption{Uncoded BER versus input back-off for the three separate soundings ($E_b/N_0=24$~dB).}
\label{fig:ibo}
\end{figure}

Fig.~\ref{fig:ibo} shows the BER versus back-off for the three separate soundings. With impulse sounding (a), LMMSE suffers from the full gain error of Table~\ref{tab:mu}. At 3~dB, its BER of $\bGL\times10^{-2}$ falls to $\bGMu\times10^{-2}$ ($-\rGMu\%$) with the single factor $1/\hat\mu$, to $\bGBA\times10^{-2}$ ($-\rGBA\%$) when the Bussgang-aware equalizer also tracks the CPE, and to $\bGTap\times10^{-2}$ ($-\rGTap\%$) with the blind one-tap tracker. LR reaches $\bGLR\times10^{-2}$ ($-\rGLR\%$), a substantial gain that is easy to attribute to learning, but it stays above all three training-free receivers at every back-off. Giving LR the same CPE tracking (LR+CPE, $\bGLRC\times10^{-2}$, $-\rGLRC\%$) brings it level with the blind one-tap tracker, which needs no training, but not with BA+CPE. With constant-envelope pilots (b) the gain error is smaller, and the improvement of LR shrinks to \rCELR\%. With data-like pilots (c), $\mu=1$, gain correction does nothing, and LR is \rDLLR\% \emph{worse} than LMMSE at 3~dB, while CPE tracking still helps (BA+CPE and the blind one-tap tracker reach $\bDLBA\times10^{-2}$ and $\bDLTap\times10^{-2}$). Blind one-tap tracking, which needs no PA model, stays within \tapOverBAmaxSixG\%, \tapOverBAmaxSixCE\%, and \tapOverBAmaxSixDL\% of BA+CPE (impulse, constant-envelope, and data-like pilots) at every back-off up to 6~dB. DD-DC shows what full PA knowledge can add.

\subsection{Embedded pilots}
\label{sec:ep}

In the AFDM embedded-pilot scheme~\cite{bemani2023afdm,yin2022pilot}, which follows the embedded pilot of OTFS~\cite{raviteja2019embedded}, one DAFT-domain pilot symbol, surrounded by guard (null) symbols, is transmitted in every frame, and the channel is read from the pilot response of the same frame. It is implemented here for the reference channel with the pilot at position 32 and the two guard positions on which the path offsets $\delta=\{0,25\}$ would otherwise place data on the pilot response (3 of 64 symbols; as for the other pilots, this overhead is not charged against $E_b/N_0$). Path delays and Dopplers are assumed known, so only the two path gains are estimated, from one noisy frame that carries its own PA distortion and phase noise. The pilot energy is boosted by 0--10~dB relative to a data symbol, and the frame power is normalized so that the back-off keeps its meaning. The pilot is added to the data \emph{before} the PA, so both are compressed together, and $\mu$ is computed by Monte Carlo on the actual frame structure (Supplementary Section~\Sref{sec:epsetup}).

\begin{table}[htbp]\centering\small
\caption{Embedded single pilot: $|\mu|$ and uncoded BER ($\times10^{-2}$).}
\label{tab:ep}
\footnotesize
\setlength{\tabcolsep}{2.2pt}
\begin{tabular*}{\tblwidth}{@{\extracolsep{\fill}}cccccccc@{}}
\toprule
Boost (dB) & $|\mu|$ & LMMSE & LMMSE/$\hat\mu$ & BA+CPE & 1-tap & LR & LR+CPE\\
\midrule
0 & 0.999 & 8.61 & 8.61 & 7.96 & 7.46 & 10.5 & 10.1 \\
3 & 0.998 & 4.73 & 4.73 & 4.30 & 4.00 & 5.60 & 5.19 \\
6 & 0.995 & 2.58 & 2.57 & 2.31 & 2.15 & 3.02 & 2.72 \\
10 & 0.987 & 1.40 & 1.33 & 1.23 & 1.09 & 1.65 & 1.53 \\

\bottomrule
\end{tabular*}
\tblnote{Boost: pilot energy relative to a data symbol. 3~dB back-off, $E_b/N_0=24$~dB, 100 runs. Other back-offs are given in Supplementary Table~\Sref{tab:epfull}.}
\end{table}

The mismatch has vanished: $|\mu|$ lies between \epMuMin{} and \epMuMax{} for every boost (0--10~dB) and back-off (1, 3, 5, and 8~dB) tested, because pilot and data sit in the same compressed waveform. The consequences follow Section~\ref{sec:theory}. Gain correction changes the BER by at most \epMuChgMax\%, and the trained LR is \epLRMin--\epLRMax\% \emph{worse} than LMMSE in every configuration (Table~\ref{tab:ep} and Supplementary Table~\Sref{tab:epfull}). What limits this scheme is the single-frame channel estimate: the PA distortion that lands on the pilot response has $\epSdTwoZero$ of the power of a data symbol, that is, a relative estimation error of about $\epErrZero$ at 0~dB boost and $\epErrTen$ at 10~dB, which is why boosting the pilot helps. The part of this error that is common to all paths acts as a per-frame gain and phase, and the blind one-tap tracker lowers the BER by \epTapMin--\epTapMax\% and BA+CPE by \epBAMin--\epBAMax\% relative to LMMSE. For an embedded pilot, then, the gain mismatch is not the issue, and the useful receiver is again the one that tracks a per-frame scalar. The results with separate soundings (Fig.~\ref{fig:ibo} and Tables~\ref{tab:robust}--\ref{tab:coded}) therefore apply to systems that measure the channel in a separate, lower-power or constant-envelope sounding (for example a power-controlled sounding reference signal or a preamble sent at a different drive level), not to a single in-frame pilot.

\subsection{Robustness}
\label{sec:robust}

\begin{table}[htbp]\centering\small
\caption{Uncoded BER ($\times10^{-2}$) with impulse sounding, changing one parameter at a time.}
\label{tab:robust}
\begin{tabular*}{\tblwidth}{@{\extracolsep{\fill}}lcccccc@{}}
\toprule
Case & LMMSE & LMMSE/$\hat\mu$ & BA+CPE & 1-tap & LR & LR+CPE\\
\midrule
Reference & 4.18 & 2.53 & 1.66 & 1.90 & 2.82 & 1.90 \\
$N=256$ & 8.93 & 7.47 & 6.44 & 7.00 & 8.19 & 7.33 \\
$N=1024$ & 18.6 & 17.9 & 17.7 & 18.1 & 19.7 & 19.7 \\
Linewidth 1~kHz & 18.6 & 18.0 & 17.7 & 18.1 & 19.6 & 19.6 \\
PN at receiver & 4.39 & 2.72 & 1.86 & 2.11 & 3.02 & 2.13 \\
PN at Tx and Rx & 7.16 & 5.57 & 4.59 & 5.08 & 6.14 & 5.26 \\
Rapp $p=1$ & 8.98 & 3.00 & 2.18 & 3.06 & 3.34 & 2.48 \\
Rapp $p=5$ & 3.08 & 2.36 & 1.49 & 1.62 & 2.64 & 1.71 \\
AM/PM 0.3~rad & 5.63 & 2.69 & 1.84 & 2.77 & 3.00 & 2.10 \\
Frac. Doppler $\nu_2{=}2.5$ & 4.19 & 2.55 & 1.68 & 1.93 & 2.84 & 1.93 \\
Random 2--4 paths & 4.42 & 2.73 & 1.88 & 2.16 & 3.06 & 2.16 \\
2 pilot frames & 4.30 & 2.61 & 1.74 & 2.01 & 3.57 & 2.42 \\
32 pilot frames & 4.25 & 2.62 & 1.76 & 2.02 & 2.78 & 1.93 \\
Model IBO off by $+2$~dB & 4.18 & 2.85 & 2.00 & 1.90 & 2.82 & 1.90 \\
Model AM/PM $0.3$ (true 0.1) & 4.18 & 3.17 & 1.86 & 1.90 & 2.82 & 1.90 \\

\bottomrule
\end{tabular*}
\tblnote[\footnotesize]{3~dB back-off, $E_b/N_0=24$~dB, and 100 runs with identical seeds in every row, except random channels with 2--4 paths and $|\nu|\le3$ (200 runs) and $N=1024$ (12 runs). The last two rows use a mismatched PA model in the receiver.}
\end{table}

Table~\ref{tab:robust} varies one parameter at a time; Supplementary Tables~\Sref{tab:robA}--\Sref{tab:robC} give the results for all three soundings (impulse and data-like pilots only for $N=256$ and 1024), and Supplementary Table~\Sref{tab:pamis} expands the two mismatched-model rows (impulse sounding). In all thirteen simulated rows (every row except the two model-mismatch rows at the bottom) LMMSE$/\hat\mu$ lowers the BER of LMMSE, BA+CPE beats both the trained LR and LR+CPE, and the blind one-tap tracker beats LR. BA+CPE removes 28--76\% of the LMMSE errors except in the two strongest phase-noise rows discussed below. The one-tap tracker ties or beats LR+CPE except where the PA is very soft ($p=1$) or has strong AM/PM (0.3~rad), where the LMMSE output carries the largest gain or phase error and a readout trained on known symbols does better than the decision-directed tracker (LR+CPE 2.48 versus 3.06 and 2.10 versus 2.77), and with 32 pilot frames (1.93 versus 2.02); training-free tracking is thus competitive with LR+CPE but not uniformly better. With two pilot frames LR loses much of its advantage (3.57), while the training-free receivers barely change. Only the rows with the strongest phase noise, a 1~kHz linewidth and $N=1024$ (a longer frame lets the phase drift further within it), depart from this picture: the error floor there is dominated by phase noise that none of these receivers tracks inside a frame, every method gains at most \robBAminAll\%, and LR is 5--6\% \emph{worse} than LMMSE. The two mismatched-model rows show that BA+CPE needs the PA model only coarsely: a 2~dB error in the back-off raises its BER from 1.66 to 2.00, and a 0.2~rad error in the AM/PM parameter from 1.66 to 1.86, still far below LMMSE (4.18). With the 2~dB back-off error, however, BA+CPE falls behind LR+CPE and the blind one-tap tracker (both 1.90).

\textit{3GPP TDL channels.} Table~\ref{tab:tdl} repeats the comparison on the TDL-C and TDL-D models~\cite{3gpp38901} as implemented in Sionna~\cite{hoydis2022sionna}, for five deployments, with the AFDM parameters scaled to each subcarrier spacing. With impulse sounding LR improves on LMMSE by \tdlLRGLo--\tdlLRGHi\%, but gain correction alone gives \tdlMuGLo--\tdlMuGHi\% and BA+CPE \tdlBAGLo--\tdlBAGHi\%; with data-like pilots LR is \tdlLRDLLo--\tdlLRDLHi\% worse than LMMSE. A one-sided paired Wilcoxon signed-rank test over 30 runs rejects the hypothesis ``LR at least as good as LMMSE/$\hat\mu$'' with $p<\tdlPmax$ in every configuration (impulse and constant-envelope sounding), and LR beats LMMSE/$\hat\mu$ in at most \tdlFracMax\% of the runs; 95\% bootstrap intervals for every entry are given in Supplementary Table~\Sref{tab:tdlci}. Training LR on one channel realization and applying it to another costs it little ($2.71\to2.83\times10^{-2}$ in configuration A), which is consistent with LR learning the PA gain rather than the channel; this test uses a separate set of runs, reported in Supplementary Section~\Sref{sec:crossreal}.

\begin{table}[htbp]\centering\small
\caption{BER change (\%) relative to LMMSE on 3GPP TDL channels.}
\label{tab:tdl}
\footnotesize
\setlength{\tabcolsep}{2.5pt}
\begin{tabular*}{\tblwidth}{@{\extracolsep{\fill}}lcccccc@{}}
\toprule
& \multicolumn{3}{c}{Impulse sounding (G)} & \multicolumn{3}{c}{Data-like pilots (DL)}\\
\cmidrule(lr){2-4}\cmidrule(lr){5-7}
Config. & LR & LMMSE/$\hat\mu$ & BA+CPE & LR & LMMSE/$\hat\mu$ & BA+CPE\\
\midrule
A & $-$31.5 & $-$38.8 & $-$58.6 & +14.4 & 0 & $-$38.6 \\
B & $-$31.1 & $-$38.5 & $-$58.9 & +14.7 & 0 & $-$39.1 \\
C & $-$59.2 & $-$63.7 & $-$71.4 & +11.8 & 0 & $-$15.4 \\
D & $-$29.3 & $-$35.9 & $-$53.6 & +13.3 & 0 & $-$32.1 \\
E & $-$34.8 & $-$41.5 & $-$64.9 & +17.0 & 0 & $-$48.4 \\

\bottomrule
\end{tabular*}
\tblnote{Negative values mean a lower BER than LMMSE. 3~dB back-off, $E_b/N_0=24$~dB, 30 runs. A: TDL-C, 3.5~GHz, 30~kHz, 120~km/h. B: as A, 500~km/h. C: TDL-C, 28~GHz, 120~kHz, 300~km/h. D: TDL-C, 28~GHz, 30~kHz, 500~km/h (maximum Doppler 0.43 subcarriers). E: TDL-D (line of sight), 3.5~GHz, 30~kHz, 120~km/h.}
\end{table}

\subsection{Coded performance and LLR calibration}
\label{sec:coded}

\begin{figure}[t]
\centering
\includegraphics[width=\linewidth]{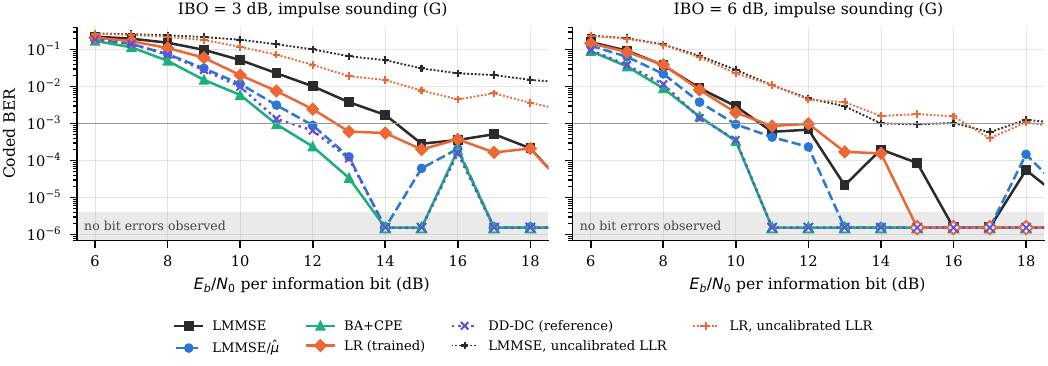}
\caption{LDPC-coded BER with impulse sounding; hollow markers: no bit errors observed.}
\label{fig:coded}
\end{figure}

\begin{table}[htbp]\centering\small
\caption{$E_b/N_0$ (dB) needed for a coded BER of $10^{-3}$.}
\label{tab:coded}
\footnotesize
\setlength{\tabcolsep}{2.5pt}
\begin{tabular*}{\tblwidth}{@{\extracolsep{\fill}}lcccccc@{}}
\toprule
Case & LMMSE & LMMSE/$\hat\mu$ & BA+CPE & 1-tap & LR & DD-DC\\
\midrule
IBO 3~dB, G & 14.3 & 11.9 & \textbf{11.0} & 12.7 & 12.6 & 11.4 \\
IBO 3~dB, DL & 12.4 & 12.4 & \textbf{12.0} & \textbf{12.0} & 13.2 & 12.2 \\
IBO 6~dB, G & 10.7 & 10.0 & \textbf{9.3} & 10.3 & 10.8 & \textbf{9.3} \\
IBO 6~dB, DL & 10.7 & 10.7 & 10.1 & 10.4 & 11.5 & \textbf{10.0} \\

\bottomrule
\end{tabular*}
\tblnote{Calibrated LLRs; best entries (within 0.05~dB) in bold. Impairment-free reference: \codedIdeal~dB. Bootstrap intervals are given in Supplementary Table~\Sref{tab:codedci}; differences quoted in the text are computed before rounding.}
\end{table}

Fig.~\ref{fig:coded} and Table~\ref{tab:coded} show the coded results. Brackets give 95\% bootstrap intervals of the difference in required $E_b/N_0$, taken as the first-named minus the second-named receiver, so negative values mean that the first receiver needs less. At 3~dB back-off with impulse sounding, LR gains \cdaThreeGLRL~dB over LMMSE \cdciThreeGLRL, gain correction alone \cdaThreeGMuL~dB \cdciThreeGMuL, and BA+CPE \cdaThreeGBAL~dB \cdciThreeGBAL; BA+CPE needs \cdaThreeGBALR~dB less than LR \cdciThreeGBALR, while the blind one-tap tracker and LR are indistinguishable (\cdThreeGTapLR~dB \cdciThreeGTapLR). At 6~dB back-off LR has no detectable coded gain: it needs \cdaSixGLRL~dB \cdciSixGLRL{} (impulse sounding) and \cdaSixDLLRL~dB \cdciSixDLLRL{} (data-like pilots) more than LMMSE, whereas BA+CPE gains \cdaSixGBAL~dB \cdciSixGBAL{} and \cdaSixDLBAL~dB \cdciSixDLBAL. With data-like pilots BA+CPE also needs \cdaThreeDLBALR~dB \cdciThreeDLBALR{} and \cdaSixDLBALR~dB \cdciSixDLBALR{} less than LR at 3 and 6~dB back-off, but only \cdaThreeDLBAL~dB \cdciThreeDLBAL{} less than LMMSE at 3~dB. In coded BER, DD-DC is not better than BA+CPE at 3~dB back-off (0.4~dB worse with impulse sounding and 0.2~dB worse with data-like pilots; Supplementary Tables~\Sref{tab:codedci} and~\Sref{tab:fer3}), so it is an upper reference only for the uncoded BER. With 1000 codewords per point and flat curves near $10^{-3}$, the intervals are coarse (Supplementary Table~\Sref{tab:codeddiff}).

The curves labelled ``uncalibrated LLR'' in Fig.~\ref{fig:coded} show what happens when the LLRs are scaled by $N_0/2$, half the complex noise variance at the receiver input. This scaling ignores noise enhancement, phase noise, and distortion, and for the max-log metric used here it overstates the reliabilities by a factor of two even without impairments, so the decoder treats distorted symbols as reliable ones. At 6~dB back-off LMMSE needs \codedUncLmmse~dB instead of \codedCalLmmse~dB, a \cdSixGUncL~dB \cdciSixGUncL{} loss caused by the LLR scaling alone, and LR needs \codedUncLR~dB instead of \codedCalLR~dB (\cdSixGUncLR~dB); at 3~dB back-off with impulse sounding, LMMSE with uncalibrated LLRs never reaches $10^{-3}$ in the sweep up to 24~dB (beyond the axis of Fig.~\ref{fig:coded}; Supplementary Table~\Sref{tab:fer3}). The uncalibrated curves flatten just above $10^{-3}$, and on such a floor a small difference in BER moves the $10^{-3}$ crossing by several decibels. The same effect appears in an independently written coded-link simulation that uses $N_0/2$ scaling and true channel knowledge (Supplementary Section~\Sref{sec:indep}): there LMMSE and LR differ by only 10--21\% in BER, yet their $10^{-3}$ crossings differ by about 4~dB, in favour of LR, whereas in the main simulation the uncalibrated comparison goes the other way. The sign and size of an apparent coded gain then depend on where the floor happens to lie. With a mismatched decoder metric~\cite{martinez2009bicm,nguyen2011bicm}, a coded comparison measures LLR scaling as much as equalization quality, so the reliabilities must be scaled by the residual variance measured by each receiver. At 3~dB back-off, isolated codeword errors at 15--18~dB appear for every receiver, including DD-DC, and are probably caused by rare large phase-noise excursions.

\subsection{Learned correctors after LMMSE}

When the 2D-RC of~\cite{xu2024twodrc} (reservoir size 12, window $4\times4$) is placed after LMMSE and trained on the same 8 pilot frames (100 runs, 32 data frames), it reaches $\rcGRC\times10^{-2}$ with impulse sounding and $\rcDLRC\times10^{-2}$ with data-like pilots. This is somewhat worse than LR (by 12--14\%; $\rcGLR\times10^{-2}$ and $\rcDLLR\times10^{-2}$) and clearly worse than LMMSE/$\hat\mu$ ($\rcGMu\times10^{-2}$, $\rcDLMu\times10^{-2}$) and BA+CPE ($\rcGBA\times10^{-2}$, $\rcDLBA\times10^{-2}$). Once LMMSE has removed the channel, the remaining problem is a gain and a phase per frame. The one reservoir configuration tested shows no advantage for this problem; Proposition~\ref{prop:readout} explains why for linear readouts.

% =======================================================
\section{Discussion and design guidelines}
\label{sec:discussion}
% =======================================================

The results suggest four rules for AFDM receivers with impaired hardware. (i)~\emph{The pilot-induced gain mismatch should be corrected or avoided.} When the channel is sounded separately with an amplitude distribution different from that of the data, the equalized data should be corrected by $\hat\mu=B_\mathrm{d}/B_\mathrm{s}$ (Table~\ref{tab:mu}); alternatively, data-like or embedded pilots make $\mu\approx1$ (Section~\ref{sec:ep}), although a single embedded pilot adds its own estimation error. (ii)~\emph{The common phase and gain should be tracked per frame.} After gain correction the dominant residual is per-frame phase noise, and with an embedded pilot it is the per-frame error of the single-frame channel estimate. A blind one-tap estimate removes a substantial part of the LMMSE errors (about \rGTap\% with separate impulse sounding, \epTapMin--\epTapMax\% with an embedded pilot) at 100~Hz linewidth and $N=64$, but not when phase drift within a frame dominates (1~kHz, $N=1024$); it adds about $0.6N^2$ multiplications per frame and needs neither training nor a PA model (BA+CPE adds the PA model). (iii)~\emph{What can be computed should not be learned.} A trained post-LMMSE readout learns a shrunk version of the same scalar, with pilot overhead and a scale bias that a QAM detector pays for; with an embedded pilot it is worse than LMMSE. (iv)~\emph{LLRs should be scaled by the measured residual variance.} At 6~dB back-off the \cdSixGUncL~dB coded penalty of mismatched LLRs for LMMSE is larger than every receiver difference observed there.

\textit{Limitations.} The SNR is referenced to the linear PA output, so the efficiency benefit of a small back-off is not credited. The PA is memoryless, the link is single-antenna, and the impulse sounding is noise-free; all results are simulations. The headline gains apply to separate soundings. The embedded-pilot results are uncoded and use the reference channel with known path delays and Dopplers, and a wider guard would be needed with fractional Doppler. The channel is assumed unchanged between sounding and data, so channel ageing, which matters for vehicular links, is not modelled, and the coded results rest on 1000 codewords per point, which resolves differences of about 1~dB or more. Nonlinear learned correctors with constellation-dependent features were not tested, only the linear readout and one 2D-RC configuration. PA memory, MIMO-AFDM, and over-the-air validation are natural next steps. The gain-mismatch argument is not specific to AFDM and should apply to any waveform with Gaussian-like samples, such as OFDM, OTFS, or orthogonal chirp division multiplexing (OCDM)~\cite{ouyang2016ocdm}.

% =======================================================
\section{Conclusion}
\label{sec:conclusion}
% =======================================================

This paper asked a simple question: when a small trained corrector is placed after the equalizer of an AFDM receiver with a nonlinear power amplifier and a noisy oscillator, what does it learn? It mostly learns one number. When the channel is measured with a pilot signal whose amplitude statistics differ from those of the data, the amplifier treats the two differently, and every equalized data symbol carries the same gain and phase error. The trained corrector learns an inverse of this error, slightly shrunk by the residual error (Proposition~\ref{prop:readout}). The same error can be computed from a model of the amplifier, or estimated from each received frame, with no training at all. Model-based gain correction with phase tracking had a lower uncoded BER than the trained corrector in every separate-sounding scenario tested with a correct PA model, and at a coded BER of $10^{-3}$ it needed \cdaThreeGBALR~dB less $E_b/N_0$ than the trained corrector with impulse sounding at 3~dB back-off, and \cdaThreeDLBALR--\cdaSixDLBALR~dB less with data-like pilots. When a pilot is embedded in every data frame, the mismatch disappears: the trained readout alone is then \epLRMin--\epLRMax\% worse than LMMSE and, even with phase tracking added, it trails the blind one-tap tracker.

Two further analytical results complete the picture. An exact identity states when dividing by the true gain lowers the mean-squared error (Proposition~\ref{prop:threshold}). For two-path channels, the inverse of the AFDM channel generally spreads over the whole frame (in the reference channel, 55\% of its energy lies outside an 8-symbol window), so short-window correctors cannot replace the equalizer (Proposition~\ref{prop:mismatch}).

For receiver designers the practical message is to correct, or avoid, the pilot-induced gain mismatch, to track one gain and one phase per frame, and to scale decoder reliabilities by the measured residual error. For researchers, the message is that a learned component should be compared against the simple model-based correction that it may be imitating. Extensions to PA memory, multiple antennas, channel ageing, and over-the-air measurements remain open.

% -------------------------------------------------------
\section*{Supplementary material}
The Supplementary Material, which follows the references, contains the complete simulation set-up, the Bussgang and pilot-gain computations, full proofs and validations of Propositions~\ref{prop:readout}--\ref{prop:mismatch}, all additional numerical results with confidence intervals and frame-error counts, and reproducibility information (Sections~\Sref{sec:sim}--\Sref{sec:repro}).

\section*{CRediT authorship contribution statement}
\textbf{Mohammad Tasin:} Conceptualization, Methodology, Software, Formal analysis, Investigation, Visualization, Writing -- original draft. \textbf{Muhammad Abdullah Arafat:} Supervision, Writing -- review \& editing.

\section*{Declaration of competing interest}
The authors declare that they have no known competing financial interests or personal relationships that could have appeared to influence the work reported in this paper.

\section*{Funding}
This research did not receive any specific grant from funding agencies in the public, commercial, or not-for-profit sectors.

\section*{Data availability}
The simulation code and the per-run results that support the findings of this study are available from the corresponding author upon reasonable request.

\section*{Declaration of generative AI and AI-assisted technologies in the manuscript preparation process}
During the preparation of this work the authors took the help of Claude Sonnet 5.5 in order to write the codes, draft and edit the text of the manuscript and of the Supplementary Material. After using this tool, the authors reviewed and edited the content as needed and take full responsibility for the content of the published article.

\bibliographystyle{elsarticle-num}
\bibliography{references}

\clearpage
\setcounter{section}{0}\setcounter{table}{0}\setcounter{equation}{0}\setcounter{figure}{0}
\renewcommand{\thesection}{S\arabic{section}}
\renewcommand{\thetable}{S\arabic{table}}
\renewcommand{\thefigure}{S\arabic{figure}}
\renewcommand{\theequation}{S\arabic{equation}}
\renewcommand{\theHsection}{S\arabic{section}}
\renewcommand{\theHtable}{S\arabic{table}}
\renewcommand{\theHfigure}{S\arabic{figure}}
\renewcommand{\theHequation}{S\arabic{equation}}
\begin{center}
{\LARGE\bfseries Supplementary Material}
\end{center}
\begin{quote}\small
This Supplementary Material accompanies the main text of this paper. Section~\ref{sec:sim} gives the complete simulation set-up, Section~\ref{sec:buss} the Bussgang and pilot-gain computations, Sections~\ref{sec:p1}--\ref{sec:p3} the full proofs and validations of Propositions~\Mref{prop:readout}--\Mref{prop:mismatch} of the main text, Section~\ref{sec:extra} all additional numerical results (including the embedded-pilot study, confidence intervals, and frame-error counts), and Section~\ref{sec:repro} the reproducibility information. All uncoded bit error rate (BER) values are averages over independent runs with common seeds, in which every receiver sees identical frames (paired comparison). Most numbers in the tables and text are generated by script from the stored per-run results (Section~\ref{sec:repro}). Sections, tables, and equations of this part carry the prefix S; numbers without the prefix refer to the main text.
\end{quote}

% ============================================================================
\section{Simulation set-up}
\label{sec:sim}
% ============================================================================

\subsection{Parameters}

Table~\ref{tab:params} lists the default parameters. Unless stated otherwise, every result uses these values.

\begin{table}[htbp]\centering\small
\caption{Default simulation parameters.}
\label{tab:params}
\begin{tabular*}{\tblwidth}{@{\extracolsep{\fill}}lp{0.62\textwidth}@{}}
\toprule
Parameter & Value\\
\midrule
Chirp subcarriers $N$, CPP length & 64, 16 samples\\
Chirp parameters $c_1$, $c_2$ & $9/128$ ($\nu_\mathrm{max}=3$), $1/128$\\
Sampling rate / spacing & 1.92~MHz / 30~kHz\\
Modulation & Gray-coded 16-QAM, unit energy\\
Reference channel & $h=[1,\,0.7+0.3j]$, $l=[0,3]$, $\nu=[0,2]$; $\delta=[0,25]$\\
PA & Rapp, $p=2$, $\alpha_\mathrm{PM}=0.1$~rad, $A_\mathrm{sat}=1$, IBO 1--8~dB (default 3~dB)\\
Phase noise & Wiener, 100~Hz linewidth, transmit oscillator after the PA\\
Pilot frames for the trained readout & 8 (random 16-QAM)\\
Data frames per run & 64 (uncoded; 16 for $N=256$, 8 for $N=1024$); three frames per codeword (coded)\\
Independent runs per point & 100 (uncoded, seeds 0--99; random channels 200, $N=1024$ 12), 30 (TDL), 10 runs $\times$ 100 codewords (coded), 100 (Proposition~\Mref{prop:threshold})\\
Embedded pilot (EP) & position 32 of 64, guards at 7 and 57 ($=32\pm25$), so 3 of 64 symbols carry no data and 61 carry data\\
Linear readout (LR) & $N_w=8$ causal cyclic window, ridge parameter $10^{-4}\mathrm{tr}(\mathbf{U}^H\mathbf{U})/N_w$\\
LDPC code & 5G NR, base graph 2, $Z=44$, $k=324$, $n=648$, 20 belief-propagation iterations (Sionna 1.2.2~\cite{hoydis2022sionna})\\
\bottomrule
\end{tabular*}
\end{table}

\subsection{Chirp-periodic prefix and channel}
The chirp-periodic prefix (CPP) repeats the last $L_\mathrm{CPP}$ samples of the block with the chirp phase correction $e^{-j2\pi c_1(N^2+2Nn)}$, $n=-L_\mathrm{CPP},\dots,-1$, which keeps the received block circular for delays up to $L_\mathrm{CPP}$~\cite{bemani2023afdm}. The multipath channel acts on the CPP-extended frame as $r[n]=\sum_i h_i e^{j2\pi\nu_i n/N}s[n-l_i]$. With this sign of the Doppler phase, path $i$ occupies the cyclic diagonal $\delta_i=(2Nc_1l_i-\nu_i)_N$ of $\Heff$; for the reference channel, $\delta_2=27-2=25$. This was verified numerically for the single paths $(l,\nu)=(0,2)$, $(3,0)$, $(3,2)$, and $(1,-1)$, which give diagonals 62, 27, 25, and 10. The convention $e^{-j2\pi\nu n/N}$ of Bemani \textit{et al.}~\cite{bemani2023afdm} gives $\delta_i=(2Nc_1l_i+\nu_i)_N$.

\subsection{SNR definition}
$E_b/N_0$ is defined per information bit, $N_0=1/(4R\cdot10^{E_b/N_0/10})$ for 16-QAM, with $R=1$ uncoded and $R=1/2$ coded. The transmit power reference is the nominal linear PA output: the PA input is scaled by $10^{-\mathrm{IBO}/20}$ and its output by $10^{+\mathrm{IBO}/20}$, so the small-signal gain is one at every input back-off (IBO). Comparisons across IBO therefore isolate the distortion and do not credit the efficiency gain of a smaller back-off.

\subsection{Pilot (sounding) schemes}
\begin{itemize}
\item \textbf{G (impulse sounding).} $N$ frames are sent, frame $k$ carrying a unit DAFT impulse at position $k$; the time samples have constant amplitude $1/\sqrt{N}$, so the PA operates in its linear region. There is no thermal noise, but each frame sees its own phase-noise path. Column $k$ of $\hat{\mathbf{H}}_\mathrm{eff}$ is the DAFT of received frame $k$.
\item \textbf{CE (constant envelope at data power).} The same impulses are scaled by $\sqrt{N}$ (time samples of unit modulus), repeated $K_\mathrm{p}/N=4$ times ($K_\mathrm{p}=256$ frames) with noise, and averaged.
\item \textbf{DL (data-like).} $K_\mathrm{p}=256$ random 16-QAM frames pass through the PA and the impaired channel; $\hat{\mathbf{H}}_\mathrm{eff}$ is the least-squares solution of $\mathbf{Y}=\hat{\mathbf{H}}_\mathrm{eff}\mathbf{X}$.
\end{itemize}

\subsection{Receivers}
\begin{itemize}
\item \textbf{LMMSE:} $\mathbf{W}=(\hat{\mathbf{H}}^H\hat{\mathbf{H}}+N_0\mathbf{I})^{-1}\hat{\mathbf{H}}^H$.
\item \textbf{LMMSE/$\hat\mu$:} $\mathbf{z}/\hat\mu$, with $\hat\mu=B_\mathrm{d}/B_\mathrm{s}$ from the PA model (Section~\ref{sec:buss}).
\item \textbf{Bussgang-aware LMMSE with CPE tracking (BA+CPE):} with $\mathbf{H}_0=\hat{\mathbf{H}}/B_\mathrm{s}$,
\begin{equation}
\mathbf{W}_\mathrm{BA}=B_\mathrm{d}^*\mathbf{H}_0^H\big((|B_\mathrm{d}|^2+\sigma_d^2)\mathbf{H}_0\mathbf{H}_0^H+N_0\mathbf{I}\big)^{-1};
\end{equation}
the output is divided element-wise by $\mathrm{diag}(\mathbf{W}_\mathrm{BA}B_\mathrm{d}\mathbf{H}_0)$, followed by two iterations of $\hat\phi=\angle\sum_m z[m]\check X^*[m]$ ($\check X[m]$: hard 16-QAM decision on $z[m]$), $\mathbf{z}\leftarrow\mathbf{z}e^{-j\hat\phi}$ per frame.
\item \textbf{Blind one-tap:} two iterations of $\hat g=\sum_m z[m]\check X^*[m]/\sum_m|\check X[m]|^2$, $\mathbf{z}\leftarrow\mathbf{z}/\hat g$ per frame, applied to the LMMSE output.
\item \textbf{Linear readout (LR):} features $\mathbf{u}[m]=[z[(m-N_w+1)_N],\dots,z[m]]$; weights $\mathbf{w}=(\mathbf{U}^H\mathbf{U}+\lambda\mathbf{I})^{-1}\mathbf{U}^H\mathbf{x}_\mathrm{p}$ over all pilot symbols, where the rows of $\mathbf{U}$ are the feature vectors and $\mathbf{x}_\mathrm{p}$ stacks the known pilot symbols; output $\mathbf{u}[m]^T\mathbf{w}$. LR+CPE adds the CPE step of BA+CPE.
\item \textbf{Decision-directed distortion cancellation (DD-DC):} initial decisions come from BA+CPE. For two iterations, $\hat{\mathbf{s}}$ is regenerated (inverse DAFT and CPP), $\hat{\mathbf{d}}=f(\hat{\mathbf{s}})-B_\mathrm{d}\hat{\mathbf{s}}$ is computed with the exact PA, $\mathbf{H}_0\,\mathrm{DAFT}(\hat{\mathbf{d}})$ is subtracted from $\mathbf{y}$, and the result is equalized with $\mathbf{W}_c=B_\mathrm{d}^*\mathbf{H}_0^H(|B_\mathrm{d}|^2\mathbf{H}_0\mathbf{H}_0^H+N_0\mathbf{I})^{-1}$, normalized, and phase-corrected with the CPE step.
\end{itemize}

\subsection{Embedded-pilot scheme}
\label{sec:epsetup}
One DAFT-domain pilot symbol at position $m_p=32$ and two guard (null) symbols at $m_p\pm25$ (positions 7 and 57) are transmitted in every frame; the guard positions are the only positions from which data would land, through one of the two paths ($\delta=\{0,25\}$), on a bin of the pilot response. The other 61 positions carry random 16-QAM symbols. The pilot has power $P_b$ relative to a data symbol ($P_b=0$, 3, 6, and 10~dB), and the frame is normalized to unit average power before the PA, whose input is then scaled by the back-off as before. Path delays and Dopplers are known; the receiver reads the two complex path gains $\hat h_i=y[(m_p-\delta_i)_N]/(A_i[(m_p-\delta_i)_N,m_p]X_p)$ from the \emph{same} noisy frame (with its own PA distortion and phase-noise path) and rebuilds $\hat{\mathbf H}_\mathrm{eff}=\sum_i\hat h_i\mathbf A_i$. The Bussgang gains of the pilot component, $B_\mathrm{s}$, and of the data component, $B_\mathrm{d}$, the distortion variance, and hence $\mu=B_\mathrm{d}/B_\mathrm{s}$ are computed by Monte Carlo simulation (4000 frames) on this frame structure. The receivers are those of the previous subsection restricted to the 61 data positions; ``true channel'' denotes LMMSE with the true channel scaled by $B_\mathrm{d}$. As for the other soundings, the overhead of the 3 non-data symbols is not charged to $E_b/N_0$.

\subsection{Coded link and LLRs}
Each 648-bit codeword is padded with 120 random bits and mapped onto three 64-symbol frames. Max-log log-likelihood ratios (LLRs) are computed as $L_b=(\min_{a\in\mathcal{A}_b^0}|\hat X-a|^2-\min_{a\in\mathcal{A}_b^1}|\hat X-a|^2)/\sigma^2$, where $\mathcal{A}_b^0$ and $\mathcal{A}_b^1$ are the constellation points whose bit $b$ is 0 and 1. In the calibrated case $\sigma^2$ is the residual variance $\E|\hat X-X|^2$ that the receiver measures on its own 8 pilot frames. In the uncalibrated case $\sigma^2=N_0/2$, half the complex noise variance at the receiver input; this ignores noise enhancement, phase noise, and distortion, and for this max-log metric it overstates the reliabilities by a factor of two even without impairments. The impairment-free reference uses an IBO of 60~dB, zero linewidth, and impulse sounding.

% ============================================================================
\section{Bussgang gains and the pilot-induced mismatch}
\label{sec:buss}
% ============================================================================

\subsection{Data gain}
For a circularly symmetric Gaussian input $x\sim\mathcal{CN}(0,1)$ (the AFDM samples are sums of $N$ chirps and close to Gaussian), $B_\mathrm{d}=\E[f(x)x^*]$ and $\sigma_d^2=\E|f(x)-B_\mathrm{d}x|^2$. With $r=|x|$ Rayleigh distributed and the Rapp amplitude response $G_\mathrm{R}(\cdot)$ and AM/PM response $\varphi(\cdot)$ of Section~\Mref{sec:system} of the main text,
\begin{equation}
B_\mathrm{d}=\int_0^\infty 2r e^{-r^2}\, r\, G_\mathrm{R}(\kappa r)e^{j\varphi(\kappa r)}\,r\,dr,\qquad \kappa=10^{-\mathrm{IBO}/20},
\end{equation}
which is evaluated by Monte Carlo simulation with $4\times10^5$ samples (relative accuracy better than $10^{-3}$). The AM/PM term rotates the gain by $\angle B_\mathrm{d}=2.4^\circ$ at 3~dB back-off.

\subsection{Pilot gains}
For a constant-envelope pilot of amplitude $a$, every sample sees the same gain, $B_\mathrm{s}=f(a)/a$. Impulse sounding has $a=1/\sqrt{N}$ and a scaled input $\kappa a\ll1$, so $B_\mathrm{s}=1+O(\kappa^{2p}a^{2p})$ with a phase of $\alpha_\mathrm{PM}\kappa^2a^2\approx0.05^\circ$. Constant-envelope pilots at data power have $a=1$ and $B_\mathrm{s}=G_\mathrm{R}(\kappa)e^{j\varphi(\kappa)}$. Data-like pilots have $B_\mathrm{s}=B_\mathrm{d}$. Table~\Mref{tab:mu} of the main text lists the values.

\subsection{From \texorpdfstring{$\mu$}{mu} to the post-equalizer gain}
With $\hat{\mathbf{H}}=B_\mathrm{s}\Heff$ (ignoring estimation noise) and $\mathbf{y}=B_\mathrm{d}\Heff\mathbf{X}+\mathbf{e}'$,
\begin{equation}
\mathbf{W}\mathbf{y}=\mu\,\big(\Heff^H\Heff+N_0|B_\mathrm{s}|^{-2}\mathbf{I}\big)^{-1}\Heff^H\Heff\,\mathbf{X}+\mathbf{W}\mathbf{e}'.
\end{equation}
The average diagonal gain is therefore $\mu$ times the usual LMMSE shrinkage factor, which tends to one at high SNR, times the average phase-noise phasor of the data relative to the sounding. At 3~dB IBO and 24~dB the measured ratio is $|g|/|\mu|=0.986$.

% ============================================================================
\section{Proposition~\Mref{prop:readout}: what a trained readout learns}
\label{sec:p1}
% ============================================================================

\subsection{Proof}
\label{sec:p1proof}
Let $z[m]=gX[m]+e[m]$ with i.i.d.\ zero-mean unit-energy $X$, $\E[e[m]X^*[m']]=0$ for all $m,m'$, $e$ zero-mean and white with variance $|g|^2v$, and $g$ the same for all $m$. The $N_w$-tap least-squares readout minimizes $\E|X[m]-\sum_j w_j z[m-j]|^2$. Its normal equations are $\sum_i w_i R_{ij}=r_j$ with $R_{ij}=\E[z[m-i]z^*[m-j]]=|g|^2(1+v)\delta_{ij}$ and $r_j=\E[X[m]z^*[m-j]]=g^*\delta_{j0}$ (Kronecker delta), so $w_j=\delta_{j0}\,g^*/(|g|^2(1+v))=\delta_{j0}/(g(1+v))$. The ridge term used here, $\lambda=10^{-4}\,\mathrm{tr}(\mathbf{U}^H\mathbf{U})/N_w$, only scales the limit by $1/(1+10^{-4})$. The output is $X/(1+v)+e/(g(1+v))$, whose MSE is $(v/(1+v))^2+v/(1+v)^2=v/(1+v)$.

\subsection{Validation}
At 3~dB IBO and 24~dB with impulse sounding (100 runs), $|g|=\poneEightEightg$ and $v=\poneEightEightv$, so the predicted value is $|w_\infty g|=1/(1+v)=\poneEightEightp$. Fitted on 8 pilot frames, the readout gives $|w_0g|=\poneEightEightw$ for $N_w=8$ and $\poneEightOnew$ for $N_w=1$, with \poneEightEighto\% of the weight energy on the seven other taps for $N_w=8$. Fitted on 256 frames, it gives $\poneTwoFiveSixEightw$ ($N_w=8$) and $\poneTwoFiveSixOnew$ ($N_w=1$). The window length has no practically relevant effect on the BER (Section~\ref{sec:robsweeps}, block ``LR window $N_w$'': the LR BER lies between \nwMin\ and \nwMax$\times10^{-2}$ for $N_w\in\{1,3,5,8,15\}$).

\subsection{Why a shrunk gain costs BER}
For 16-QAM with decision thresholds at $0$ and $\pm2/\sqrt{10}$, an output $\beta X+n$ with $\beta<1$ moves the outer points ($\pm3/\sqrt{10}$) towards the thresholds by $3(1-\beta)/\sqrt{10}$ per dimension, while the inner points move towards the zero threshold by only $(1-\beta)/\sqrt{10}$. The MSE-optimal $\beta=1/(1+v)$ therefore trades a small noise reduction for a systematic loss on 50\% of the symbol coordinates. The measured BER at 3~dB and 24~dB is $\bGMu\times10^{-2}$ for $z/\hat\mu$ and $\bGLR\times10^{-2}$ for LR.

% ============================================================================
\section{Proposition~\Mref{prop:threshold}: exact correction criterion}
\label{sec:p2}
% ============================================================================

\subsection{Proof and finite-pilot extension}
\label{sec:p2proof}
The identity $\mathrm{MSE}(z)-\mathrm{MSE}(z/g)=|1-g|^2-v(1-|g|^2)$ follows from $\mathrm{MSE}(z)=|1-g|^2+|g|^2v$ and $\mathrm{MSE}(z/g)=v$ (proof of Proposition~\Mref{prop:threshold} in the main text). It is algebraic: evaluated on each run with that run's own $g$ and $v$, it holds to $\pTwoAlg$ (maximum over all runs and settings below). If $g$ is replaced by an estimate $\hat g=g(1+\epsilon)$ with $\E[\epsilon]=0$ and $\E|\epsilon|^2=s^2$, independent of the data frame, then $\mathrm{MSE}(z/\hat g)\approx v+s^2(1+v)$, and the criterion becomes
\begin{equation}
|1-g|^2>v(1-|g|^2)+s^2(1+v).
\label{eq:finitepilot}
\end{equation}
For a scalar estimated from $M$ symbols that share the phase of the data frame (for example, decision-directed over the frame itself), $s^2\approx v/M$, so the extra term $s^2(1+v)$ is about $1\times10^{-3}$ for the $M=64$ symbols of one frame, which is comparable to the identity values near the crossover (Fig.~\Mref{fig:threshold}(a) of the main text). For estimates from \emph{other} frames, $s^2$ also contains the frame-to-frame variation of the gain caused by the common phase error, which does not decrease with $M$. This is the dominant penalty of a readout trained on separate pilot frames.

\subsection{Residual-variance budget}

\begin{table}[htbp]\centering\small
\caption{Unbiased residual variance $v$ at 24~dB with impulse sounding, with the impairments switched on one at a time (100 runs).}
\label{tab:budget}
\begin{tabular*}{\tblwidth}{@{\extracolsep{\fill}}lccccc@{}}
\toprule
IBO & Noise only & + PN (100~Hz) & + PA only & PA + PN & Sum of increments\\
\midrule
3~dB & 0.0024 & 0.0372 & 0.0227 & 0.0583 & 0.0575 \\
5.5~dB & 0.0024 & 0.0372 & 0.0090 & 0.0434 & 0.0438 \\
8~dB & 0.0024 & 0.0372 & 0.0040 & 0.0396 & 0.0388 \\

\bottomrule
\end{tabular*}
\end{table}

Table~\ref{tab:budget} shows that the sum of the separate increments is within \budgetGapNum\% of the joint value at every IBO. The phase-noise (PN) contribution ($\approx\budgetPN$ above the noise floor) does not depend on the IBO and dominates for IBO $\geq3$~dB, so the crossover of Proposition~\Mref{prop:threshold} is set mainly by the oscillator.

\subsection{Full validation table}
Table~\ref{tab:p2} lists the measurements behind Fig.~\Mref{fig:threshold} of the main text (100 runs per point, impulse sounding). The columns are $E_b/N_0$ (dB), IBO (dB), $|g|$ and $v$ (run averages), the identity evaluated at these averages (``plug-in'', $\times10^{-3}$), the measured MSE gains of $z/g$ and of LR over LMMSE (run averages, $\times10^{-3}$), and the BER ($\times10^{-2}$) of LMMSE, $z/g$, and LR. The run-averaged measured gain of $z/g$ equals the run average of the per-run identity (to $\pTwoAlg$); the plug-in column differs from it by a Jensen gap of at most $\pTwoJensen\times10^{-3}$ because $|g|$ and $v$ vary from run to run.

\begin{table}[htbp]\centering\small
\caption{Validation of Proposition~\Mref{prop:threshold} (impulse sounding, 100 runs per point).}
\label{tab:p2}
\footnotesize
\begin{tabular*}{\tblwidth}{@{\extracolsep{\fill}}cccccccccc@{}}
\toprule
 & & & & \multicolumn{3}{c}{MSE gain over LMMSE ($\times10^{-3}$)} & \multicolumn{3}{c}{BER ($\times10^{-2}$)}\\
\cmidrule(lr){5-7}\cmidrule(lr){8-10}
$E_b/N_0$ & IBO & $|g|$ & $v$ & Plug-in & $z/g$ & LR & LMMSE & $z/g$ & LR\\
\midrule
18 & 4.0 & 0.866 & 0.0597 & +4.16 & +4.35 & +5.37 & 3.77 & 2.68 & 3.01 \\
18 & 4.5 & 0.881 & 0.0565 & +2.62 & +2.95 & +3.32 & 3.29 & 2.44 & 2.81 \\
18 & 5.0 & 0.895 & 0.0540 & +1.55 & +1.82 & +2.17 & 2.95 & 2.25 & 2.58 \\
18 & 5.5 & 0.907 & 0.0518 & +0.29 & +0.61 & +1.10 & 2.63 & 2.11 & 2.40 \\
18 & 6.0 & 0.918 & 0.0503 & $-$0.45 & $-$0.10 & +0.08 & 2.41 & 2.01 & 2.32 \\
18 & 6.5 & 0.928 & 0.0488 & $-$0.87 & $-$0.52 & $-$0.93 & 2.25 & 1.92 & 2.24 \\
18 & 7.0 & 0.936 & 0.0480 & $-$1.24 & $-$0.95 & $-$1.51 & 2.14 & 1.89 & 2.22 \\
18 & 7.5 & 0.943 & 0.0462 & $-$1.26 & $-$0.93 & $-$1.20 & 1.97 & 1.74 & 2.02 \\
18 & 8.0 & 0.949 & 0.0466 & $-$1.68 & $-$1.36 & $-$1.63 & 1.95 & 1.79 & 2.06 \\
24 & 4.0 & 0.872 & 0.0507 & +5.49 & +5.70 & +5.68 & 2.99 & 2.04 & 2.36 \\
24 & 4.5 & 0.888 & 0.0483 & +3.60 & +3.96 & +3.80 & 2.63 & 1.90 & 2.19 \\
24 & 5.0 & 0.901 & 0.0461 & +2.21 & +2.51 & +2.11 & 2.31 & 1.75 & 2.05 \\
24 & 5.5 & 0.914 & 0.0439 & +1.13 & +1.42 & +1.35 & 2.06 & 1.63 & 1.87 \\
24 & 6.0 & 0.924 & 0.0426 & +0.35 & +0.73 & +0.32 & 1.90 & 1.56 & 1.80 \\
24 & 6.5 & 0.934 & 0.0409 & $-$0.18 & +0.19 & $-$0.51 & 1.73 & 1.46 & 1.71 \\
24 & 7.0 & 0.943 & 0.0394 & $-$0.54 & $-$0.22 & $-$0.94 & 1.57 & 1.38 & 1.61 \\
24 & 7.5 & 0.950 & 0.0395 & $-$0.93 & $-$0.64 & $-$1.37 & 1.56 & 1.40 & 1.63 \\
24 & 8.0 & 0.956 & 0.0385 & $-$0.99 & $-$0.57 & $-$0.84 & 1.49 & 1.35 & 1.53 \\
30 & 4.0 & 0.874 & 0.0480 & +5.39 & +5.69 & +5.48 & 2.75 & 1.87 & 2.17 \\
30 & 4.5 & 0.889 & 0.0462 & +3.71 & +4.05 & +3.63 & 2.47 & 1.79 & 2.06 \\
30 & 5.0 & 0.903 & 0.0437 & +2.41 & +2.70 & +2.56 & 2.15 & 1.63 & 1.85 \\
30 & 5.5 & 0.915 & 0.0419 & +1.38 & +1.73 & +1.22 & 1.94 & 1.53 & 1.77 \\
30 & 6.0 & 0.926 & 0.0400 & +0.57 & +0.92 & +0.62 & 1.74 & 1.43 & 1.63 \\
30 & 6.5 & 0.936 & 0.0392 & $-$0.20 & +0.13 & $-$0.67 & 1.62 & 1.39 & 1.62 \\
30 & 7.0 & 0.944 & 0.0382 & $-$0.44 & $-$0.10 & $-$0.91 & 1.54 & 1.34 & 1.57 \\
30 & 7.5 & 0.951 & 0.0383 & $-$0.92 & $-$0.58 & $-$0.84 & 1.51 & 1.36 & 1.55 \\
30 & 8.0 & 0.958 & 0.0372 & $-$0.90 & $-$0.60 & $-$1.73 & 1.42 & 1.30 & 1.55 \\

\bottomrule
\end{tabular*}
\end{table}

The identity (run-averaged) changes sign between an IBO of $\pTwoCrossLo$ and $\pTwoCrossHi$~dB over the three SNRs; the measured MSE crossover of LR is at $\pTwoLRLo$--$\pTwoLRHi$~dB, and its BER crossover at $\pTwoBerLo$--$\pTwoBerHi$~dB (linear interpolation of the table). The BER of $z/g$ is below that of LMMSE at \pTwoBerAll{} point.

% ============================================================================
\section{Proposition~\Mref{prop:mismatch}: structure of the inverse AFDM channel}
\label{sec:p3}
% ============================================================================

\subsection{Proof details}
\label{sec:p3proof}
Let $\mathbf{A}_i$ denote the monomial matrix of path $i$ (unit-modulus entries on cyclic diagonal $\delta_i$). Then $\mathbf{M}=\mathbf{A}_1^{-1}\mathbf{A}_2$ is monomial on diagonal $\Delta s=(\delta_2-\delta_1)_N$, with $[\mathbf{M}]_{m,(m+\Delta s)_N}=\gamma_m$ and $|\gamma_m|=1$. Hence $[\mathbf{M}^k]_{m,(m+k\Delta s)_N}=\prod_{t=0}^{k-1}\gamma_{(m+t\Delta s)_N}$, and $\mathbf{M}^N$ is diagonal with entries $\prod_{t=0}^{N-1}\gamma_{(m+t\Delta s)_N}$ (every orbit of length $N/\gcd(\Delta s,N)$ is traversed $\gcd(\Delta s,N)$ times). Numerically, for the reference channel all diagonal entries of $\mathbf{M}^{64}$ are equal and have unit modulus to machine precision (about $10^{-15}$). The truncated sum $\sum_{k<N}(-c)^k\mathbf{M}^k$ matches $(\mathbf{I}+c\mathbf{M})^{-1}$ to a relative error of $2.7\times10^{-8}$, in agreement with $|c|^{64}=0.58^{32}=2.7\times10^{-8}$.

If $\gcd(\Delta s,N)=d>1$, the $d$ terms $k, k+n,\dots,k+(d-1)n$ with $n=N/d$ fall on the same diagonal and their energies add; Eq.~\Mref{eq:rho} of the main text then applies with $\omega$ summed over each diagonal. Precisely,
\begin{equation}
(\mathbf{I}+c\mathbf{M})^{-1}=\sum_{k<n}(-c)^k\mathbf{M}^k\,(\mathbf{I}-(-c)^n\mathbf{M}^n)^{-1};
\end{equation}
row $m$ of diagonal $k$ has modulus $|c|^k/|1-(-c)^n\lambda_m|$ with $|\lambda_m|=1$, so its energy fraction is $(1-|c|^2)|c|^{2k}/(1-|c|^{2n})=\sum_{j=0}^{d-1}\omega_{k+jn}$. For $\Delta s=N/2$, for example, $\Heff^{-1}$ has only two diagonals (0 and $N/2$), and any window that excludes diagonal $N/2$ misses a fraction $|c|^2/(1+|c|^2)$ of the energy.

\subsection{Geometry sweep}
A total of 168 two-path geometries ($l_2\in\{1,\dots,8\}$, $\nu_2\in\{-3,\dots,3\}$, $|c|\in\{0.5,0.76,0.9\}$, $\angle c=0.6$~rad) were evaluated. The empirical diagonal of $\mathbf{M}$ equalled $(9l_2-\nu_2)_{64}$ in all cases, and Eq.~\Mref{eq:rho} of the main text for the symmetric band $|d|\le7$ matched the numerical projection to $6\times10^{-14}$. Table~\ref{tab:geo} summarizes the mismatch.

\begin{table}[htbp]\centering\small
\caption{Relative Frobenius mismatch $\rho_\mathcal{B}$ of the symmetric $\pm7$ band (15 taps) over 168 geometries: mean [min, max]. The numbers in parentheses give the geometries per value of $|c|$ for which $\Delta s$ lies outside or inside the band.}
\label{tab:geo}
\begin{tabular*}{\tblwidth}{@{\extracolsep{\fill}}lccc@{}}
\toprule
$|c|$ & $\Delta s$ outside band (43) & $\Delta s$ inside band (13) & $\mathrm{cond}(\Heff)$\\
\midrule
0.50 & 0.492 [0.447, 0.500] & 0.154 [0.000, 0.250] & 2.9--3.0\\
0.76 & 0.725 [0.612, 0.757] & 0.412 [0.000, 0.577] & 6.1--7.3\\
0.90 & 0.830 [0.740, 0.876] & 0.642 [0.000, 0.797] & 9.4--18.9\\
\bottomrule
\end{tabular*}
\end{table}

For the reference channel, the condition number is $\mathrm{cond}(\Heff)=7.381$, against the bound $(1+|c|)/(1-|c|)=7.39$; $\Heff$ has 36.7\% of its energy off the main diagonal, and $\Heff^{-1}$ has 58.0\%. The quoted mismatch is a Frobenius-norm ratio: $\rho=\rhoCausal$ for the causal 8-lag window (diagonals $0,-1,\dots,-7$), $\rhoSym$ for the symmetric $\pm7$ band, and $\rhoFwd$ for a forward-only 8-lag window; the corresponding energy fraction is $\rhoCausal^2\approx\rhoEnergy\%$.

\subsection{Causal and windowed equalizers}
For a support set $\mathcal{S}_m$ of observations allowed for output $m$, the MMSE-optimal linear estimator is $\hat X[m]=\mathbf{h}_{\mathcal{S}_m,m}^H\mathbf{R}_{\mathcal{S}_m}^{-1}\mathbf{y}_{\mathcal{S}_m}$, with $\mathbf{R}=\Heff\Heff^H+N_0\mathbf{I}$ and $\mathbf{h}_{\cdot,m}$ the $m$-th column of $\Heff$. It was evaluated at 24~dB with 4000 noise-only frames (no PA and no phase noise); Table~\ref{tab:causal} gives the results.

\begin{table}[htbp]\centering\small
\caption{BER and MSE of the best linear estimator with restricted support (reference channel, 24~dB, no hardware impairments).}
\label{tab:causal}
\begin{tabular*}{\tblwidth}{@{\extracolsep{\fill}}lcc@{}}
\toprule
Support & BER & MSE\\
\midrule
Full (LMMSE) & 0.0000 & 0.0023\\
Causal, unlimited memory ($n\le m$) & 0.160 & 0.230\\
Anti-causal ($n\ge m$) & 0.153 & 0.214\\
Causal cyclic window, $N_w=8$ & 0.253 & 0.368\\
Cyclic window $N_w=16$ / 24 / 26--32 & 0.253 / 0.253 / 0.218 & --\\
Cyclic window $N_w=40$ / 48 & 0.064 & 0.089\\
Cyclic window $N_w=56$ / 64 & 0.000 & --\\
\bottomrule
\end{tabular*}
\end{table}

A forward-running reservoir with a causal window implements, in its linear regime, a lower-triangular map whose memory decays as $\rho_\mathrm{sr}^k$, where $\rho_\mathrm{sr}$ is the spectral radius ($0.9^{25}=0.07$, $0.9^{50}=0.005$). The value 0.160 is therefore a benchmark for linear maps, not a bound for a trained reservoir (nor a guarantee that a nonlinear reservoir cannot do better).

\subsection{Readouts on the strongest diagonals and on time-domain windows}
Two limits of the structural statement of Proposition~\Mref{prop:mismatch} are quantified here (reference channel, 24~dB, noise only). \emph{Offset-aware readouts:} the best linear estimator that uses, for output $m$, the $K$ observations on the $K$ strongest diagonals of $\Heff^{-1}$ (offsets known to the receiver). \emph{Time-domain windows:} the best linear estimator that uses only $W$ consecutive \emph{time-domain} samples (the inverse DAFT spreads each symbol over all $N$ samples with equal magnitude). Table~\ref{tab:offtime} lists the mismatch $\rho_\mathcal{B}$, MSE, and BER.

\begin{table}[htbp]\centering\small
\caption{Offset-aware readout on the $K$ strongest diagonals of $\Heff^{-1}$ (left) and the best linear estimator on $W$ consecutive time-domain samples (right); reference channel, 24~dB, noise only.}
\label{tab:offtime}
\begin{tabular}{@{}cccc@{}}
\toprule
\multicolumn{4}{c}{Offset-aware, $K$ strongest diagonals}\\
$K$ & $\rho_\mathcal{B}$ & MSE & BER\\
\midrule
1 & 0.762 & -- & -- \\
2 & 0.580 & 0.1767 & 0.144 \\
4 & 0.336 & 0.0525 & 0.022 \\
8 & 0.113 & 0.0076 & 0.000 \\
12 & 0.038 & 0.0029 & 0.000 \\
16 & 0.013 & 0.0024 & 0.000 \\

\bottomrule
\end{tabular}\qquad\qquad
\begin{tabular}{@{}ccc@{}}
\toprule
\multicolumn{3}{c}{Time-domain window}\\
$W$ & MSE & BER\\
\midrule
8 & 0.876 & 0.434 \\
16 & 0.753 & 0.392 \\
32 & 0.501 & 0.300 \\
48 & 0.252 & 0.184 \\
56 & 0.126 & 0.096 \\
60 & 0.065 & 0.036 \\
64 & 0.002 & 0.000 \\

\bottomrule
\end{tabular}
\end{table}

The offset-aware readout works (BER \offBerEight\ at $K=8$) but requires the channel geometry; it is a sparse model-based inverse rather than a generic reservoir. Time-domain windows fail like DAFT-domain windows and reach zero BER only for $W$ close to $N$.

\subsection{Reservoir detectors on raw observations}
The echo-state network (ESN) has the state $\mathbf{q}[m]=\tanh(\mathbf{W}_i\mathbf{y}_w[m]+\mathbf{W}_r\mathbf{q}[m-1])$, with input window 8, sparsity 0.6, a ridge readout on $[\mathbf{q};\mathbf{y}_w]$, and inputs normalized to unit power (IBO 3~dB with the default PA and phase noise, 24~dB, 4 runs). The 2D-RC of~\cite{xu2024twodrc} adapted to AFDM uses reservoir size 12, a window of $M_w=4$ subcarriers $\times$ 4 frames, three reservoir matrices (row, column, diagonal), spectral radius 0.9, and sparsity 0.6 (same conditions, 100 runs, 8 pilot frames, 32 data frames). Table~\ref{tab:esn} gives the results.

\begin{table}[htbp]\centering\small
\caption{BER of reservoir detectors on raw DAFT-domain observations ($N_n$: reservoir size; sr: spectral radius), and of the 2D-RC placed after LMMSE.}
\label{tab:esn}
\begin{tabular*}{\tblwidth}{@{\extracolsep{\fill}}lc@{}}
\toprule
Detector & BER\\
\midrule
ESN, $N_n=32$, sr 0.9, 8 / 64 pilot frames & 0.270 / 0.254\\
ESN, $N_n=128$, sr 0.9 / 0.99, 64 frames & 0.256 / 0.257\\
ESN, $N_n=512$, sr 0.9 / 0.99, 64 frames & 0.272 / 0.272\\
ESN, $N_n=512$, sr 0.99, 256 frames & 0.257\\
2D-RC (raw input), 8 frames & 0.267\\
2D-RC after LMMSE, G / DL sounding (100 runs, 32 data frames) & $\rcGRC\times10^{-2}$ / $\rcDLRC\times10^{-2}$\\
\bottomrule
\end{tabular*}
\end{table}

\FloatBarrier
% ============================================================================
\section{Additional numerical results}
\label{sec:extra}
% ============================================================================

\FloatBarrier
\subsection{One-at-a-time robustness sweeps}
\label{sec:robsweeps}
Tables~\ref{tab:robA}--\ref{tab:robC} extend Table~\Mref{tab:robust} of the main text to all three soundings (IBO 3~dB, 24~dB; BER $\times10^{-2}$). G denotes impulse sounding, CE constant-envelope pilots, and DL data-like pilots; blocks without CE rows were run with G and DL only. The columns give the number of runs and the BER of LMMSE, LMMSE/$\hat\mu$, BA+CPE, the blind one-tap tracker, LR, LR+CPE, and DD-DC (reference).

\begin{table}[htbp]\centering\small
\caption{Robustness sweeps, part 1: frame length, oscillator linewidth, and phase-noise placement (BER $\times10^{-2}$).}
\label{tab:robA}
\footnotesize
\begin{tabular*}{\tblwidth}{@{\extracolsep{\fill}}llrccccccc@{}}
\toprule
Case & Pilots & Runs & LMMSE & LMMSE/$\hat\mu$ & BA+CPE & 1-tap & LR & LR+CPE & DD-DC\\
\midrule
\multicolumn{10}{@{}l}{\textit{Frame length $N$}} \\
$N=64$ & G & 100 & 4.18 & 2.53 & 1.66 & 1.90 & 2.82 & 1.90 & 0.64 \\
$N=64$ & CE & 100 & 2.08 & 1.54 & 0.68 & 0.74 & 1.79 & 0.83 & 0.10 \\
$N=64$ & DL & 100 & 2.13 & 2.13 & 1.24 & 1.25 & 2.43 & 1.45 & 0.26 \\
$N=256$ & G & 100 & 8.93 & 7.47 & 6.44 & 7.00 & 8.19 & 7.33 & 5.80 \\
$N=256$ & DL & 100 & 5.83 & 5.83 & 4.07 & 4.12 & 6.58 & 4.97 & 2.85 \\
$N=1024$ & G & 12 & 18.6 & 17.9 & 17.7 & 18.1 & 19.7 & 19.7 & 18.3 \\
$N=1024$ & DL & 12 & 14.9 & 14.9 & 13.6 & 13.7 & 16.5 & 15.9 & 13.7 \\
\midrule
\multicolumn{10}{@{}l}{\textit{Oscillator linewidth}} \\
1~Hz & G & 100 & 0.96 & 0.15 & 0.15 & 0.15 & 0.17 & 0.17 & 0.00 \\
1~Hz & CE & 100 & 0.41 & 0.16 & 0.16 & 0.16 & 0.18 & 0.19 & 0.00 \\
1~Hz & DL & 100 & 0.41 & 0.41 & 0.42 & 0.42 & 0.47 & 0.48 & 0.00 \\
100~Hz (ref.) & G & 100 & 4.18 & 2.53 & 1.66 & 1.90 & 2.82 & 1.90 & 0.64 \\
100~Hz (ref.) & CE & 100 & 2.08 & 1.54 & 0.68 & 0.74 & 1.79 & 0.83 & 0.10 \\
100~Hz (ref.) & DL & 100 & 2.13 & 2.13 & 1.24 & 1.25 & 2.43 & 1.45 & 0.26 \\
1~kHz & G & 100 & 18.6 & 18.0 & 17.7 & 18.1 & 19.6 & 19.6 & 18.1 \\
1~kHz & CE & 100 & 13.6 & 13.5 & 12.3 & 12.7 & 14.9 & 14.4 & 12.0 \\
1~kHz & DL & 100 & 15.1 & 15.1 & 14.0 & 14.2 & 16.7 & 16.3 & 14.1 \\
10~kHz & G & 100 & 44.8 & 45.0 & 44.8 & 44.7 & 47.6 & 47.6 & 45.0 \\
10~kHz & CE & 100 & 43.4 & 43.5 & 43.4 & 42.9 & 46.0 & 46.0 & 43.6 \\
10~kHz & DL & 100 & 45.7 & 45.7 & 45.8 & 45.3 & 46.9 & 46.9 & 45.8 \\
\midrule
\multicolumn{10}{@{}l}{\textit{Phase-noise placement}} \\
Tx (ref.) & G & 100 & 4.18 & 2.53 & 1.66 & 1.90 & 2.82 & 1.90 & 0.64 \\
Tx (ref.) & CE & 100 & 2.08 & 1.54 & 0.68 & 0.74 & 1.79 & 0.83 & 0.10 \\
Tx (ref.) & DL & 100 & 2.13 & 2.13 & 1.24 & 1.25 & 2.43 & 1.45 & 0.26 \\
Rx & G & 100 & 4.39 & 2.72 & 1.86 & 2.11 & 3.02 & 2.13 & 0.77 \\
Rx & CE & 100 & 2.19 & 1.64 & 0.77 & 0.83 & 1.90 & 0.93 & 0.12 \\
Rx & DL & 100 & 2.26 & 2.26 & 1.36 & 1.37 & 2.56 & 1.59 & 0.33 \\
Tx and Rx & G & 100 & 7.16 & 5.57 & 4.59 & 5.08 & 6.14 & 5.26 & 3.67 \\
Tx and Rx & CE & 100 & 4.08 & 3.50 & 1.97 & 2.19 & 3.98 & 2.47 & 0.96 \\
Tx and Rx & DL & 100 & 4.28 & 4.28 & 2.81 & 2.87 & 4.86 & 3.43 & 1.58 \\

\bottomrule
\end{tabular*}
\end{table}

\begin{table}[htbp]\centering\small
\caption{Robustness sweeps, part 2: PA smoothness, AM/PM, Doppler of the second path, and random channels (BER $\times10^{-2}$).}
\label{tab:robB}
\footnotesize
\begin{tabular*}{\tblwidth}{@{\extracolsep{\fill}}llrccccccc@{}}
\toprule
Case & Pilots & Runs & LMMSE & LMMSE/$\hat\mu$ & BA+CPE & 1-tap & LR & LR+CPE & DD-DC\\
\midrule
\multicolumn{10}{@{}l}{\textit{Rapp smoothness $p$}} \\
$p=1$ & G & 100 & 8.98 & 3.00 & 2.18 & 3.06 & 3.34 & 2.48 & 0.73 \\
$p=1$ & CE & 100 & 2.54 & 2.00 & 1.10 & 1.17 & 2.29 & 1.31 & 0.12 \\
$p=1$ & DL & 100 & 2.81 & 2.81 & 1.91 & 1.93 & 3.17 & 2.21 & 0.42 \\
$p=2$ (ref.) & G & 100 & 4.18 & 2.53 & 1.66 & 1.90 & 2.82 & 1.90 & 0.64 \\
$p=2$ (ref.) & CE & 100 & 2.08 & 1.54 & 0.68 & 0.74 & 1.79 & 0.83 & 0.10 \\
$p=2$ (ref.) & DL & 100 & 2.13 & 2.13 & 1.24 & 1.25 & 2.43 & 1.45 & 0.26 \\
$p=3$ & G & 100 & 3.41 & 2.41 & 1.54 & 1.70 & 2.69 & 1.76 & 0.63 \\
$p=3$ & CE & 100 & 1.93 & 1.43 & 0.60 & 0.65 & 1.67 & 0.73 & 0.10 \\
$p=3$ & DL & 100 & 1.96 & 1.96 & 1.08 & 1.09 & 2.24 & 1.27 & 0.24 \\
$p=5$ & G & 100 & 3.08 & 2.36 & 1.49 & 1.62 & 2.64 & 1.71 & 0.65 \\
$p=5$ & CE & 100 & 1.81 & 1.39 & 0.57 & 0.61 & 1.62 & 0.70 & 0.11 \\
$p=5$ & DL & 100 & 1.89 & 1.89 & 1.03 & 1.03 & 2.18 & 1.20 & 0.24 \\
\midrule
\multicolumn{10}{@{}l}{\textit{AM/PM $\alpha_{\mathrm{PM}}$}} \\
0 & G & 100 & 4.00 & 2.51 & 1.64 & 1.81 & 2.80 & 1.88 & 0.63 \\
0 & CE & 100 & 2.05 & 1.52 & 0.67 & 0.72 & 1.77 & 0.81 & 0.10 \\
0 & DL & 100 & 2.10 & 2.10 & 1.21 & 1.22 & 2.40 & 1.41 & 0.27 \\
0.1~rad (ref.) & G & 100 & 4.18 & 2.53 & 1.66 & 1.90 & 2.82 & 1.90 & 0.64 \\
0.1~rad (ref.) & CE & 100 & 2.08 & 1.54 & 0.68 & 0.74 & 1.79 & 0.83 & 0.10 \\
0.1~rad (ref.) & DL & 100 & 2.13 & 2.13 & 1.24 & 1.25 & 2.43 & 1.45 & 0.26 \\
0.3~rad & G & 100 & 5.63 & 2.69 & 1.84 & 2.77 & 3.00 & 2.10 & 0.66 \\
0.3~rad & CE & 100 & 2.31 & 1.70 & 0.82 & 0.90 & 1.96 & 0.98 & 0.11 \\
0.3~rad & DL & 100 & 2.37 & 2.37 & 1.47 & 1.48 & 2.69 & 1.72 & 0.30 \\
\midrule
\multicolumn{10}{@{}l}{\textit{Doppler of path 2}} \\
$\nu_2=2$ (ref.) & G & 100 & 4.18 & 2.53 & 1.66 & 1.90 & 2.82 & 1.90 & 0.64 \\
$\nu_2=2$ (ref.) & CE & 100 & 2.08 & 1.54 & 0.68 & 0.74 & 1.79 & 0.83 & 0.10 \\
$\nu_2=2$ (ref.) & DL & 100 & 2.13 & 2.13 & 1.24 & 1.25 & 2.43 & 1.45 & 0.26 \\
2.25 & G & 100 & 4.20 & 2.54 & 1.67 & 1.92 & 2.84 & 1.92 & 0.64 \\
2.25 & CE & 100 & 2.07 & 1.54 & 0.69 & 0.74 & 1.80 & 0.83 & 0.10 \\
2.25 & DL & 100 & 2.14 & 2.14 & 1.25 & 1.25 & 2.44 & 1.45 & 0.27 \\
2.5 & G & 100 & 4.19 & 2.55 & 1.68 & 1.93 & 2.84 & 1.93 & 0.63 \\
2.5 & CE & 100 & 2.07 & 1.54 & 0.68 & 0.74 & 1.79 & 0.82 & 0.10 \\
2.5 & DL & 100 & 2.13 & 2.13 & 1.24 & 1.25 & 2.44 & 1.45 & 0.26 \\
2.75 & G & 100 & 4.19 & 2.56 & 1.67 & 1.93 & 2.85 & 1.92 & 0.64 \\
2.75 & CE & 100 & 2.09 & 1.54 & 0.69 & 0.74 & 1.80 & 0.82 & 0.10 \\
2.75 & DL & 100 & 2.15 & 2.15 & 1.24 & 1.25 & 2.45 & 1.46 & 0.26 \\
\midrule
\multicolumn{10}{@{}l}{\textit{Random channels}} \\
2--4 paths & G & 200 & 4.42 & 2.73 & 1.88 & 2.16 & 3.06 & 2.16 & 0.90 \\
2--4 paths & CE & 200 & 2.33 & 1.76 & 0.93 & 1.00 & 2.03 & 1.11 & 0.33 \\
2--4 paths & DL & 200 & 2.36 & 2.36 & 1.49 & 1.50 & 2.68 & 1.73 & 0.57 \\

\bottomrule
\end{tabular*}
\end{table}

\begin{table}[htbp]\centering\small
\caption{Robustness sweeps, part 3: number of pilot frames and readout window (BER $\times10^{-2}$).}
\label{tab:robC}
\footnotesize
\begin{tabular*}{\tblwidth}{@{\extracolsep{\fill}}llrccccccc@{}}
\toprule
Case & Pilots & Runs & LMMSE & LMMSE/$\hat\mu$ & BA+CPE & 1-tap & LR & LR+CPE & DD-DC\\
\midrule
\multicolumn{10}{@{}l}{\textit{Pilot frames for LR}} \\
2 & G & 100 & 4.30 & 2.61 & 1.74 & 2.01 & 3.57 & 2.42 & 0.69 \\
2 & CE & 100 & 2.09 & 1.55 & 0.69 & 0.75 & 2.41 & 1.12 & 0.09 \\
2 & DL & 100 & 2.15 & 2.15 & 1.23 & 1.23 & 3.10 & 1.84 & 0.25 \\
4 & G & 100 & 4.31 & 2.61 & 1.74 & 2.01 & 3.10 & 2.12 & 0.69 \\
4 & CE & 100 & 2.08 & 1.55 & 0.68 & 0.73 & 1.95 & 0.90 & 0.09 \\
4 & DL & 100 & 2.14 & 2.14 & 1.24 & 1.24 & 2.62 & 1.55 & 0.26 \\
8 (ref.) & G & 100 & 4.18 & 2.53 & 1.66 & 1.90 & 2.82 & 1.90 & 0.64 \\
8 (ref.) & CE & 100 & 2.08 & 1.54 & 0.68 & 0.74 & 1.79 & 0.83 & 0.10 \\
8 (ref.) & DL & 100 & 2.13 & 2.13 & 1.24 & 1.25 & 2.43 & 1.45 & 0.26 \\
16 & G & 100 & 4.23 & 2.58 & 1.70 & 1.95 & 2.79 & 1.91 & 0.65 \\
16 & CE & 100 & 2.10 & 1.55 & 0.67 & 0.73 & 1.72 & 0.79 & 0.10 \\
16 & DL & 100 & 2.15 & 2.15 & 1.23 & 1.23 & 2.36 & 1.40 & 0.25 \\
32 & G & 100 & 4.25 & 2.62 & 1.76 & 2.02 & 2.78 & 1.93 & 0.71 \\
32 & CE & 100 & 2.08 & 1.55 & 0.69 & 0.75 & 1.67 & 0.78 & 0.10 \\
32 & DL & 100 & 2.12 & 2.12 & 1.22 & 1.22 & 2.30 & 1.37 & 0.25 \\
\midrule
\multicolumn{10}{@{}l}{\textit{LR window $N_w$}} \\
1 & G & 100 & 4.18 & 2.53 & 1.66 & 1.90 & 2.77 & 1.84 & 0.64 \\
1 & CE & 100 & 2.08 & 1.54 & 0.68 & 0.74 & 1.74 & 0.80 & 0.10 \\
1 & DL & 100 & 2.13 & 2.13 & 1.24 & 1.25 & 2.36 & 1.39 & 0.26 \\
3 & G & 100 & 4.18 & 2.53 & 1.66 & 1.90 & 2.78 & 1.87 & 0.64 \\
3 & CE & 100 & 2.08 & 1.54 & 0.68 & 0.74 & 1.76 & 0.80 & 0.10 \\
3 & DL & 100 & 2.13 & 2.13 & 1.24 & 1.25 & 2.38 & 1.41 & 0.26 \\
5 & G & 100 & 4.18 & 2.53 & 1.66 & 1.90 & 2.80 & 1.88 & 0.64 \\
5 & CE & 100 & 2.08 & 1.54 & 0.68 & 0.74 & 1.78 & 0.81 & 0.10 \\
5 & DL & 100 & 2.13 & 2.13 & 1.24 & 1.25 & 2.40 & 1.42 & 0.26 \\
8 (ref.) & G & 100 & 4.18 & 2.53 & 1.66 & 1.90 & 2.82 & 1.90 & 0.64 \\
8 (ref.) & CE & 100 & 2.08 & 1.54 & 0.68 & 0.74 & 1.79 & 0.83 & 0.10 \\
8 (ref.) & DL & 100 & 2.13 & 2.13 & 1.24 & 1.25 & 2.43 & 1.45 & 0.26 \\
15 & G & 100 & 4.18 & 2.53 & 1.66 & 1.90 & 2.89 & 1.96 & 0.64 \\
15 & CE & 100 & 2.08 & 1.54 & 0.68 & 0.74 & 1.84 & 0.86 & 0.10 \\
15 & DL & 100 & 2.13 & 2.13 & 1.24 & 1.25 & 2.49 & 1.50 & 0.26 \\

\bottomrule
\end{tabular*}
\end{table}

All rows share the seeds $0,\dots,99$ (frames, data, noise, and phase-noise draws are identical for every receiver), and the reference row appears in each block. The $N$ sweep keeps $f_s$ and the linewidth fixed, so longer frames accumulate more phase noise; the $N=256$ rows use 16 data frames per run, and the $N=1024$ rows use 12 runs of 8 data frames (LR, like the other trained and model-based receivers, uses 8 pilot frames at every $N$). The random channels have $P\in\{2,3,4\}$ paths, delays in $\{0,\dots,8\}$ with an exponential power profile, and Doppler shifts uniform in $[-3,3]$ (200 runs); LR beats LMMSE/$\hat\mu$ in \randWinG{} of \randN{} paired runs with impulse sounding and in \randWinDL{} of \randN{} with data-like pilots. With data-like pilots, LMMSE$/\hat\mu$ coincides with LMMSE because $\hat\mu=1$. At 10~kHz linewidth every receiver is at the error floor (BER $\approx0.45$), and no receiver helps.

\FloatBarrier
\subsection{PA-model mismatch}
In Table~\ref{tab:pamis} the transmitter uses the true PA (Rapp, $p=2$, $\alpha_\mathrm{PM}=0.1$~rad, IBO 3~dB), whereas the receiver computes $\hat\mu$, $B_\mathrm{d}$, and $\sigma_d^2$ from a wrong model (impulse sounding, 24~dB, 100 runs). The receivers that do not use the PA model (LMMSE, the blind one-tap tracker, LR, and LR+CPE) are unaffected, as the constant columns show.

\begin{table}[htbp]\centering\small
\caption{BER ($\times10^{-2}$) with a mismatched PA model in the receiver.}
\label{tab:pamis}
\begin{tabular*}{\tblwidth}{@{\extracolsep{\fill}}lcccccc@{}}
\toprule
PA model in the receiver & LMMSE & LMMSE/$\hat\mu$ & BA+CPE & 1-tap & LR & LR+CPE\\
\midrule
Exact PA model (ref.) & 4.18 & 2.53 & 1.66 & 1.90 & 2.82 & 1.90 \\
IBO $-1$~dB & 4.18 & 2.56 & 1.70 & 1.90 & 2.82 & 1.90 \\
IBO $+1$~dB & 4.18 & 2.64 & 1.78 & 1.90 & 2.82 & 1.90 \\
IBO $+2$~dB & 4.18 & 2.85 & 2.00 & 1.90 & 2.82 & 1.90 \\
Rapp $p=3$ (true 2) & 4.18 & 2.64 & 1.78 & 1.90 & 2.82 & 1.90 \\
Rapp $p=1.5$ & 4.18 & 2.55 & 1.69 & 1.90 & 2.82 & 1.90 \\
AM/PM 0 (true 0.1) & 4.18 & 2.68 & 1.70 & 1.90 & 2.82 & 1.90 \\
AM/PM 0.3 & 4.18 & 3.17 & 1.86 & 1.90 & 2.82 & 1.90 \\

\bottomrule
\end{tabular*}
\end{table}

\FloatBarrier
\subsection{BER versus IBO (reference channel)}
Table~\ref{tab:ibo} lists selected values behind Fig.~\Mref{fig:ibo} of the main text (100 runs per point, 24~dB).

\begin{table}[htbp]\centering\small
\caption{Uncoded BER ($\times10^{-2}$) versus input back-off for the three separate soundings.}
\label{tab:ibo}
\begin{tabular*}{\tblwidth}{@{\extracolsep{\fill}}llccccccc@{}}
\toprule
Pilots & IBO & LMMSE & LMMSE/$\hat\mu$ & BA+CPE & 1-tap & LR & LR+CPE & DD-DC\\
\midrule
G & 1~dB & 8.65 & 4.12 & 3.45 & 4.24 & 4.56 & 3.92 & 1.14 \\
G & 3~dB & 4.18 & 2.53 & 1.66 & 1.90 & 2.82 & 1.90 & 0.64 \\
G & 5~dB & 2.31 & 1.80 & 0.90 & 0.99 & 2.05 & 1.04 & 0.54 \\
G & 8~dB & 1.49 & 1.40 & 0.52 & 0.55 & 1.53 & 0.60 & 0.47 \\
CE & 1~dB & 4.23 & 3.06 & 2.23 & 2.38 & 3.45 & 2.58 & 0.20 \\
CE & 3~dB & 2.08 & 1.54 & 0.68 & 0.74 & 1.79 & 0.83 & 0.10 \\
CE & 5~dB & 1.02 & 0.85 & 0.18 & 0.20 & 1.04 & 0.24 & 0.06 \\
CE & 8~dB & 0.60 & 0.57 & 0.06 & 0.08 & 0.66 & 0.09 & 0.05 \\
DL & 1~dB & 4.38 & 4.38 & 3.67 & 3.68 & 4.86 & 4.15 & 1.20 \\
DL & 3~dB & 2.13 & 2.13 & 1.24 & 1.25 & 2.43 & 1.45 & 0.26 \\
DL & 5~dB & 1.02 & 1.02 & 0.30 & 0.30 & 1.23 & 0.38 & 0.11 \\
DL & 8~dB & 0.58 & 0.58 & 0.08 & 0.09 & 0.68 & 0.11 & 0.07 \\

\bottomrule
\end{tabular*}
\end{table}

\FloatBarrier
\subsection{Embedded pilot: full results}
Table~\ref{tab:epfull} extends Table~\Mref{tab:ep} of the main text to back-offs of 1, 3, 5, and 8~dB (reference channel, 24~dB, 100 runs; pilot boost relative to a data symbol; $|\mu|$ from the Monte Carlo model of Section~\ref{sec:epsetup}). The ``true channel'' column (LMMSE with the true channel scaled by $B_\mathrm{d}$) indicates how much of the loss comes from the single-frame channel estimate: the gap between LMMSE and this receiver is large at low boost and closes as the boost grows. It is not a bound: at 10~dB boost for back-offs of 3~dB or more, and at 6--10~dB boost for 8~dB back-off, LMMSE with the same-frame estimate is better, because that estimate also absorbs the common phase error of the frame, which the true-channel receiver lacks.

\begin{table}[htbp]\centering\small
\caption{Embedded single pilot: $|\mu|$ and uncoded BER ($\times10^{-2}$) for back-offs of 1, 3, 5, and 8~dB and all pilot boosts.}
\label{tab:epfull}
\begin{tabular*}{\tblwidth}{@{\extracolsep{\fill}}cccccccccc@{}}
\toprule
IBO (dB) & Boost (dB) & $|\mu|$ & LMMSE & LMMSE/$\hat\mu$ & BA+CPE & 1-tap & LR & LR+CPE & True ch.\\
\midrule
1 & 0 & 0.998 & 13.9 & 13.8 & 13.3 & 12.6 & 17.0 & 16.8 & 2.48 \\
1 & 3 & 0.996 & 9.15 & 9.13 & 8.59 & 8.16 & 11.3 & 11.0 & 2.56 \\
1 & 6 & 0.993 & 5.99 & 5.96 & 5.58 & 5.30 & 7.04 & 6.73 & 2.66 \\
1 & 10 & 0.982 & 4.05 & 3.91 & 3.73 & 3.53 & 4.72 & 4.57 & 3.21 \\
\midrule
3 & 0 & 0.999 & 8.61 & 8.61 & 7.96 & 7.46 & 10.5 & 10.1 & 1.11 \\
3 & 3 & 0.998 & 4.73 & 4.73 & 4.30 & 4.00 & 5.60 & 5.19 & 1.16 \\
3 & 6 & 0.995 & 2.58 & 2.57 & 2.31 & 2.15 & 3.02 & 2.72 & 1.20 \\
3 & 10 & 0.987 & 1.40 & 1.33 & 1.23 & 1.09 & 1.65 & 1.53 & 1.48 \\
\midrule
5 & 0 & 0.999 & 4.14 & 4.14 & 3.69 & 3.41 & 4.85 & 4.35 & 0.58 \\
5 & 3 & 0.999 & 1.91 & 1.90 & 1.69 & 1.57 & 2.16 & 1.89 & 0.57 \\
5 & 6 & 0.997 & 0.73 & 0.73 & 0.63 & 0.58 & 0.83 & 0.70 & 0.59 \\
5 & 10 & 0.992 & 0.34 & 0.33 & 0.30 & 0.25 & 0.39 & 0.36 & 0.81 \\
\midrule
8 & 0 & 1.000 & 1.42 & 1.42 & 1.25 & 1.17 & 1.57 & 1.36 & 0.35 \\
8 & 3 & 1.000 & 0.47 & 0.47 & 0.41 & 0.38 & 0.50 & 0.42 & 0.35 \\
8 & 6 & 0.999 & 0.16 & 0.16 & 0.14 & 0.13 & 0.17 & 0.14 & 0.36 \\
8 & 10 & 0.997 & 0.07 & 0.07 & 0.06 & 0.05 & 0.08 & 0.07 & 0.48 \\

\bottomrule
\end{tabular*}
\end{table}

\FloatBarrier
\subsection{3GPP TDL channels}
\label{sec:tdl}
The channels are generated with the TDL-C and TDL-D models of 3GPP TR 38.901~\cite{3gpp38901} as implemented in Sionna~\cite{hoydis2022sionna}, with a delay spread of 300~ns (100~ns for configuration C), taps resampled to the AFDM sampling rate, and one realization per run (30 runs). Table~\ref{tab:tdlconf} lists the configurations, Table~\ref{tab:tdlfull} the BER (IBO 3~dB, 24~dB), Table~\ref{tab:tdlci} the BER changes with bootstrap intervals, and Tables~\ref{tab:tdlibo} and~\ref{tab:tdlsnr} the dependence on back-off and SNR.

\begin{table}[htbp]\centering\small
\caption{TDL configurations ($f_DN/f_s$: maximum Doppler shift in subcarrier spacings; DS: delay spread in samples).}
\label{tab:tdlconf}
\begin{tabular*}{\tblwidth}{@{\extracolsep{\fill}}lccccccc@{}}
\toprule
Config. & Model & $f_c$ & Spacing & Speed & $f_s$ & $f_DN/f_s$ & DS (samples)\\
\midrule
A & TDL-C & 3.5~GHz & 30~kHz & 120~km/h & 1.92~MHz & 0.013 & 0.58\\
B & TDL-C & 3.5~GHz & 30~kHz & 500~km/h & 1.92~MHz & 0.054 & 0.58\\
C & TDL-C & 28~GHz & 120~kHz & 300~km/h & 7.68~MHz & 0.065 & 0.77\\
D & TDL-C & 28~GHz & 30~kHz & 500~km/h & 1.92~MHz & 0.432 & 0.58\\
E & TDL-D & 3.5~GHz & 30~kHz & 120~km/h & 1.92~MHz & 0.013 & 0.58\\
\bottomrule
\end{tabular*}
\end{table}

\begin{table}[htbp]\centering\small
\caption{Uncoded BER ($\times10^{-2}$) on the TDL channels, IBO 3~dB, 24~dB, 30 runs.}
\label{tab:tdlfull}
\begin{tabular*}{\tblwidth}{@{\extracolsep{\fill}}llcccccc@{}}
\toprule
Config. & Pilots & LMMSE & LMMSE/$\hat\mu$ & BA+CPE & 1-tap & LR & DD-DC\\
\midrule
A & G & 4.18 & 2.56 & 1.73 & 1.98 & 2.87 & 0.73 \\
A & CE & 2.16 & 1.61 & 0.80 & 0.88 & 1.89 & 0.20 \\
A & DL & 2.21 & 2.21 & 1.36 & 1.36 & 2.53 & 0.40 \\
B & G & 4.17 & 2.56 & 1.71 & 1.94 & 2.87 & 0.67 \\
B & CE & 2.16 & 1.59 & 0.76 & 0.82 & 1.87 & 0.15 \\
B & DL & 2.21 & 2.21 & 1.35 & 1.35 & 2.54 & 0.34 \\
C & G & 2.18 & 0.79 & 0.62 & 0.65 & 0.89 & 0.10 \\
C & CE & 1.16 & 0.71 & 0.53 & 0.54 & 0.80 & 0.09 \\
C & DL & 1.16 & 1.16 & 0.98 & 0.98 & 1.30 & 0.22 \\
D & G & 4.79 & 3.07 & 2.22 & 2.54 & 3.39 & 1.16 \\
D & CE & 2.61 & 2.02 & 1.16 & 1.25 & 2.29 & 0.39 \\
D & DL & 2.67 & 2.67 & 1.81 & 1.81 & 3.02 & 0.74 \\
E & G & 3.74 & 2.19 & 1.31 & 1.52 & 2.44 & 0.38 \\
E & CE & 1.74 & 1.26 & 0.45 & 0.48 & 1.49 & 0.04 \\
E & DL & 1.78 & 1.78 & 0.92 & 0.91 & 2.08 & 0.11 \\

\bottomrule
\end{tabular*}
\end{table}

\begin{table}[htbp]\centering\footnotesize
\caption{BER change relative to LMMSE on the TDL channels (\%, negative = better): mean and 95\% bootstrap interval over the 30 runs.}
\label{tab:tdlci}
\begin{tabular*}{\tblwidth}{@{\extracolsep{\fill}}llcccc@{}}
\toprule
Config. & Pilots & LR & LMMSE/$\hat\mu$ & BA+CPE & 1-tap\\
\midrule
A & G & $-$31.5 [$-$34.3, $-$28.6] & $-$38.8 [$-$41.3, $-$36.5] & $-$58.6 [$-$62.5, $-$54.8] & $-$52.7 [$-$56.8, $-$48.5] \\
A & CE & $-$12.5 [$-$15.8, $-$8.6] & $-$25.2 [$-$27.3, $-$23.3] & $-$62.9 [$-$68.4, $-$57.3] & $-$59.3 [$-$65.4, $-$53.6] \\
A & DL & +14.4 [+10.8, +18.6] & +0.0 [+0.0, +0.0] & $-$38.6 [$-$43.3, $-$34.1] & $-$38.4 [$-$43.0, $-$33.9] \\
B & G & $-$31.1 [$-$33.6, $-$28.7] & $-$38.5 [$-$40.8, $-$36.4] & $-$58.9 [$-$61.9, $-$55.7] & $-$53.6 [$-$56.9, $-$50.1] \\
B & CE & $-$13.4 [$-$16.7, $-$9.6] & $-$26.3 [$-$28.0, $-$24.6] & $-$64.8 [$-$68.5, $-$60.6] & $-$61.8 [$-$65.9, $-$57.7] \\
B & DL & +14.7 [+10.6, +19.3] & +0.0 [+0.0, +0.0] & $-$39.1 [$-$43.0, $-$35.4] & $-$38.9 [$-$42.5, $-$35.3] \\
C & G & $-$59.2 [$-$63.2, $-$55.6] & $-$63.7 [$-$67.6, $-$60.1] & $-$71.4 [$-$75.9, $-$67.2] & $-$70.4 [$-$74.7, $-$66.4] \\
C & CE & $-$31.0 [$-$35.7, $-$27.3] & $-$39.3 [$-$43.9, $-$35.6] & $-$54.0 [$-$59.4, $-$49.6] & $-$53.2 [$-$59.7, $-$47.9] \\
C & DL & +11.8 [+9.6, +14.4] & +0.0 [+0.0, +0.0] & $-$15.4 [$-$18.9, $-$12.8] & $-$15.5 [$-$19.0, $-$13.0] \\
D & G & $-$29.3 [$-$31.5, $-$27.3] & $-$35.9 [$-$38.1, $-$33.9] & $-$53.6 [$-$56.4, $-$50.9] & $-$46.9 [$-$50.4, $-$43.5] \\
D & CE & $-$12.1 [$-$15.1, $-$9.1] & $-$22.7 [$-$24.6, $-$21.0] & $-$55.3 [$-$59.8, $-$51.2] & $-$51.9 [$-$56.9, $-$47.5] \\
D & DL & +13.3 [+10.0, +17.1] & +0.0 [+0.0, +0.0] & $-$32.1 [$-$36.0, $-$28.8] & $-$32.2 [$-$35.8, $-$29.0] \\
E & G & $-$34.8 [$-$37.0, $-$32.3] & $-$41.5 [$-$43.3, $-$39.8] & $-$64.9 [$-$66.7, $-$63.1] & $-$59.5 [$-$61.3, $-$57.8] \\
E & CE & $-$14.6 [$-$18.2, $-$10.5] & $-$27.5 [$-$28.8, $-$26.2] & $-$74.4 [$-$75.9, $-$72.8] & $-$72.7 [$-$73.9, $-$71.3] \\
E & DL & +17.0 [+12.9, +21.5] & +0.0 [+0.0, +0.0] & $-$48.4 [$-$49.9, $-$46.8] & $-$48.6 [$-$50.2, $-$46.9] \\

\bottomrule
\end{tabular*}
\end{table}

\begin{table}[htbp]\centering\small
\caption{BER change relative to LMMSE (\%, negative = better) versus IBO on the TDL channels, 24~dB.}
\label{tab:tdlibo}
\begin{tabular*}{\tblwidth}{@{\extracolsep{\fill}}lllccccc@{}}
\toprule
Config. & Pilots & Receiver & 1~dB & 3~dB & 4.5~dB & 6~dB & 8~dB\\
\midrule
A & G & LR & $-$46.3 & $-$31.5 & $-$18.5 & $-$4.7 & +3.0 \\
A & G & LMMSE/$\hat\mu$ & $-$51.6 & $-$38.8 & $-$26.4 & $-$15.2 & $-$7.6 \\
A & G & BA+CPE & $-$59.6 & $-$58.6 & $-$60.1 & $-$60.6 & $-$62.7 \\
A & CE & LR & $-$17.4 & $-$12.5 & $-$6.3 & +7.7 & +9.6 \\
A & CE & LMMSE/$\hat\mu$ & $-$26.7 & $-$25.2 & $-$19.9 & $-$11.9 & $-$5.3 \\
A & CE & BA+CPE & $-$46.5 & $-$62.9 & $-$74.5 & $-$75.2 & $-$80.3 \\
A & DL & LR & +11.8 & +14.4 & +15.9 & +22.2 & +18.4 \\
A & DL & LMMSE/$\hat\mu$ & +0.0 & +0.0 & +0.0 & +0.0 & +0.0 \\
A & DL & BA+CPE & $-$16.6 & $-$38.6 & $-$60.3 & $-$67.6 & $-$75.7 \\
E & G & LR & $-$48.6 & $-$34.8 & $-$20.0 & $-$4.9 & +1.3 \\
E & G & LMMSE/$\hat\mu$ & $-$53.8 & $-$41.5 & $-$27.3 & $-$16.2 & $-$7.6 \\
E & G & BA+CPE & $-$63.3 & $-$64.9 & $-$66.3 & $-$68.9 & $-$71.4 \\
E & CE & LR & $-$18.8 & $-$14.6 & $-$5.7 & +6.9 & +13.7 \\
E & CE & LMMSE/$\hat\mu$ & $-$29.0 & $-$27.5 & $-$19.7 & $-$13.9 & $-$5.6 \\
E & CE & BA+CPE & $-$51.9 & $-$74.4 & $-$85.9 & $-$90.6 & $-$92.2 \\
E & DL & LR & +12.5 & +17.0 & +17.5 & +25.4 & +22.0 \\
E & DL & LMMSE/$\hat\mu$ & +0.0 & +0.0 & +0.0 & +0.0 & +0.0 \\
E & DL & BA+CPE & $-$19.9 & $-$48.4 & $-$73.7 & $-$85.4 & $-$88.7 \\

\bottomrule
\end{tabular*}
\end{table}

\begin{table}[htbp]\centering\small
\caption{BER ($\times10^{-2}$) versus $E_b/N_0$ on the TDL channels, IBO 3~dB.}
\label{tab:tdlsnr}
\begin{tabular*}{\tblwidth}{@{\extracolsep{\fill}}lllcccc@{}}
\toprule
Config. & Pilots & Receiver & 12~dB & 18~dB & 24~dB & 30~dB\\
\midrule
A & G & LMMSE & 8.15 & 5.28 & 4.18 & 4.10 \\
A & G & LMMSE/$\hat\mu$ & 6.07 & 3.47 & 2.56 & 2.48 \\
A & G & BA+CPE & 5.50 & 2.61 & 1.73 & 1.61 \\
A & G & 1-tap & 5.95 & 2.99 & 1.98 & 1.91 \\
A & G & LR & 6.53 & 3.93 & 2.87 & 2.80 \\
A & G & DD-DC & 4.77 & 1.48 & 0.73 & 0.64 \\
A & DL & LMMSE & 6.66 & 3.37 & 2.21 & 1.97 \\
A & DL & LMMSE/$\hat\mu$ & 6.66 & 3.37 & 2.21 & 1.97 \\
A & DL & BA+CPE & 6.04 & 2.46 & 1.36 & 1.13 \\
A & DL & 1-tap & 6.13 & 2.50 & 1.36 & 1.13 \\
A & DL & LR & 7.20 & 3.90 & 2.53 & 2.27 \\
A & DL & DD-DC & 5.51 & 1.29 & 0.40 & 0.29 \\
C & G & LMMSE & 7.90 & 3.88 & 2.18 & 1.77 \\
C & G & LMMSE/$\hat\mu$ & 5.61 & 2.02 & 0.79 & 0.54 \\
C & G & BA+CPE & 5.33 & 1.77 & 0.62 & 0.40 \\
C & G & 1-tap & 5.73 & 1.90 & 0.65 & 0.43 \\
C & G & LR & 5.99 & 2.22 & 0.89 & 0.63 \\
C & G & DD-DC & 4.64 & 0.86 & 0.10 & 0.04 \\
C & DL & LMMSE & 7.03 & 2.78 & 1.16 & 0.76 \\
C & DL & LMMSE/$\hat\mu$ & 7.03 & 2.78 & 1.16 & 0.76 \\
C & DL & BA+CPE & 6.81 & 2.55 & 0.98 & 0.59 \\
C & DL & 1-tap & 6.90 & 2.55 & 0.98 & 0.58 \\
C & DL & LR & 7.47 & 3.03 & 1.30 & 0.85 \\
C & DL & DD-DC & 6.47 & 1.54 & 0.22 & 0.06 \\

\bottomrule
\end{tabular*}
\end{table}

\FloatBarrier
\subsection{Cross-realization test}
\label{sec:crossreal}
To check whether LR learns the channel or the PA gain, LR was trained on one channel realization and tested on another, in a separate set of runs (hence the slightly different LMMSE values compared with Table~\ref{tab:tdlfull}). The BER values ($\times10^{-2}$) are, in the order LMMSE, LR trained on the same realization, LR trained on another realization, and LMMSE/$\hat\mu$: configuration A, 4.08, 2.71, 2.83, 2.46; configuration B, 4.08, 2.72, 2.82, 2.46; configuration C, 2.34, 0.99, 1.01, 0.90.

\FloatBarrier
\subsection{Coded results: confidence intervals and differences}
Table~\ref{tab:codedci} gives the $E_b/N_0$ (dB per information bit) needed for a coded BER of $10^{-3}$, with 95\% bootstrap intervals over the 10 runs of 100 codewords each. The crossing is found by linear interpolation in $\log_{10}$BER between the grid points of Section~\ref{sec:fer}; a grid point without any bit error is assigned a BER of half an error over the transmitted information bits, so that the crossing stays interpolated. A run is the resampling unit, and with 10 runs the intervals are coarse; no interval is given where more than 10\% of the resamples never reach $10^{-3}$. The impairment-free reference needs \codedIdeal~dB. Table~\ref{tab:codeddiff} gives paired differences in the required $E_b/N_0$ (first-named minus second-named receiver) with 95\% bootstrap intervals from the same resamples.

\begin{table}[htbp]\centering\small
\caption{$E_b/N_0$ (dB) needed for a coded BER of $10^{-3}$, with 95\% bootstrap intervals (calibrated LLRs unless marked ``uncal.'').}
\label{tab:codedci}
\footnotesize
\begin{tabular*}{\tblwidth}{@{\extracolsep{\fill}}llcc@{}}
\toprule
Case & Receiver / LLR & Point & 95\% interval\\
\midrule
3~dB, G & LMMSE & 14.30 & 13.9--14.8 \\
3~dB, G & LMMSE/$\hat\mu$ & 11.91 & 11.5--12.2 \\
3~dB, G & BA+CPE & 10.98 & 10.7--11.2 \\
3~dB, G & 1-tap & 12.69 & 12.2--13.1 \\
3~dB, G & LR & 12.64 & 12.2--14.0 \\
3~dB, G & LR+CPE & 12.59 & 11.9--13.1 \\
3~dB, G & DD-DC & 11.41 & 10.9--12.1 \\
\midrule
3~dB, DL & LMMSE & 12.45 & 12.0--12.7 \\
3~dB, DL & LMMSE/$\hat\mu$ & 12.45 & 12.0--12.7 \\
3~dB, DL & BA+CPE & 12.05 & 11.6--12.2 \\
3~dB, DL & 1-tap & 12.03 & 11.8--12.1 \\
3~dB, DL & LR & 13.23 & 12.7--14.3 \\
3~dB, DL & LR+CPE & 12.84 & 12.6--13.1 \\
3~dB, DL & DD-DC & 12.21 & 11.9--12.4 \\
3~dB, DL & LMMSE, uncal.\ LLR & 19.84 & ------ \\
\midrule
6~dB, G & LMMSE & 10.68 & 10.4--11.4 \\
6~dB, G & LMMSE/$\hat\mu$ & 9.97 & 9.6--10.5 \\
6~dB, G & BA+CPE & 9.29 & 8.9--9.7 \\
6~dB, G & 1-tap & 10.30 & 9.9--11.0 \\
6~dB, G & LR & 10.82 & 10.0--12.1 \\
6~dB, G & LR+CPE & 10.36 & 9.7--11.0 \\
6~dB, G & DD-DC & 9.26 & 8.9--9.6 \\
6~dB, G & LMMSE, uncal.\ LLR & 14.25 & 13.5--16.8 \\
6~dB, G & LR, uncal.\ LLR & 16.33 & 13.8--16.7 \\
\midrule
6~dB, DL & LMMSE & 10.71 & 10.2--11.1 \\
6~dB, DL & LMMSE/$\hat\mu$ & 10.71 & 10.2--11.1 \\
6~dB, DL & BA+CPE & 10.08 & 9.7--10.3 \\
6~dB, DL & 1-tap & 10.38 & 10.1--10.7 \\
6~dB, DL & LR & 11.50 & 10.7--12.6 \\
6~dB, DL & LR+CPE & 10.91 & 10.3--12.1 \\
6~dB, DL & DD-DC & 9.98 & 9.7--10.2 \\
6~dB, DL & LMMSE, uncal.\ LLR & 13.85 & 13.4--16.3 \\
6~dB, DL & LR, uncal.\ LLR & 16.36 & 15.1--17.0 \\

\bottomrule
\end{tabular*}
\end{table}

\begin{table}[htbp]\centering\small
\caption{Paired differences in the required $E_b/N_0$ (dB; negative = the first receiver needs less), with 95\% bootstrap intervals.}
\label{tab:codeddiff}
\footnotesize
\begin{tabular*}{\tblwidth}{@{\extracolsep{\fill}}llcc@{}}
\toprule
Case & Difference & Point & 95\% interval\\
\midrule
3~dB, G & LR $-$ LMMSE & $-$1.66 & [$-$2.26, $-$0.26] \\
3~dB, G & LMMSE/$\hat\mu$ $-$ LMMSE & $-$2.39 & [$-$2.89, $-$1.95] \\
3~dB, G & BA+CPE $-$ LMMSE & $-$3.33 & [$-$3.72, $-$2.86] \\
3~dB, G & BA+CPE $-$ LR & $-$1.67 & [$-$2.85, $-$1.25] \\
3~dB, G & 1-tap $-$ LR & +0.04 & [$-$1.30, +0.69] \\
3~dB, G & LR $-$ LR+CPE & +0.05 & [$-$0.32, +1.14] \\
\midrule
3~dB, DL & LR $-$ LMMSE & +0.78 & [+0.38, +1.81] \\
3~dB, DL & LMMSE/$\hat\mu$ $-$ LMMSE & +0.00 & [$-$0.00, +0.00] \\
3~dB, DL & BA+CPE $-$ LMMSE & $-$0.40 & [$-$0.78, $-$0.12] \\
3~dB, DL & BA+CPE $-$ LR & $-$1.18 & [$-$2.27, $-$0.69] \\
3~dB, DL & 1-tap $-$ LR & $-$1.20 & [$-$2.23, $-$0.69] \\
3~dB, DL & LR $-$ LR+CPE & +0.39 & [$-$0.02, +1.51] \\
\midrule
6~dB, G & LR $-$ LMMSE & +0.14 & [$-$0.60, +1.49] \\
6~dB, G & LMMSE/$\hat\mu$ $-$ LMMSE & $-$0.71 & [$-$1.36, $-$0.26] \\
6~dB, G & BA+CPE $-$ LMMSE & $-$1.39 & [$-$2.15, $-$0.90] \\
6~dB, G & BA+CPE $-$ LR & $-$1.54 & [$-$2.76, $-$1.04] \\
6~dB, G & 1-tap $-$ LR & $-$0.52 & [$-$1.81, +0.09] \\
6~dB, G & LR $-$ LR+CPE & +0.46 & [$-$0.08, +1.62] \\
6~dB, G & LMMSE, uncal.\ LLR $-$ LMMSE & +3.56 & [+2.65, +6.11] \\
6~dB, G & LR, uncal.\ LLR $-$ LR & +5.51 & [+2.37, +6.25] \\
\midrule
6~dB, DL & LR $-$ LMMSE & +0.80 & [+0.24, +1.91] \\
6~dB, DL & LMMSE/$\hat\mu$ $-$ LMMSE & +0.00 & [+0.00, +0.01] \\
6~dB, DL & BA+CPE $-$ LMMSE & $-$0.62 & [$-$1.19, $-$0.25] \\
6~dB, DL & BA+CPE $-$ LR & $-$1.42 & [$-$2.79, $-$0.65] \\
6~dB, DL & 1-tap $-$ LR & $-$1.13 & [$-$2.32, $-$0.50] \\
6~dB, DL & LR $-$ LR+CPE & +0.59 & [$-$0.27, +1.83] \\
6~dB, DL & LMMSE, uncal.\ LLR $-$ LMMSE & +3.15 & [+2.54, +5.74] \\
6~dB, DL & LR, uncal.\ LLR $-$ LR & +4.85 & [+2.95, +6.11] \\

\bottomrule
\end{tabular*}
\end{table}

At 3~dB back-off with impulse sounding, BA+CPE needs \cdaThreeGBAL~dB less than LMMSE, and LR \cdaThreeGLRL~dB less. With data-like pilots, BA+CPE needs \cdaThreeDLBAL~dB and \cdaSixDLBAL~dB less than LMMSE and \cdaThreeDLBALR~dB and \cdaSixDLBALR~dB less than LR at 3 and 6~dB back-off. At 6~dB back-off the interval for LR $-$ LMMSE with calibrated LLRs (\cdciSixGLRL~dB with impulse sounding) contains zero, that is, no coded gain of LR is detected. LLRs scaled by $N_0/2$ shift the requirement by several dB (\cdSixGUncL~dB for LMMSE and \cdSixGUncLR~dB for LR at an IBO of 6~dB with impulse sounding) and by different amounts for different receivers, so the gap between receivers changes with the LLR scaling (LR $-$ LMMSE at 6~dB with impulse sounding: $+0.1$~dB calibrated and $+2.1$~dB uncalibrated); across implementations even its sign can change (Section~\ref{sec:indep}).

\FloatBarrier
\subsection{Coded results: frame-error counts}
\label{sec:fer}
Tables~\ref{tab:fer3} and~\ref{tab:fer6} give the number of erroneous codewords out of 1000 per point (10 runs $\times$ 100 codewords). The columns are LMMSE (L), LMMSE/$\hat\mu$, BA+CPE, the blind one-tap tracker, LR, LR+CPE, DD-DC, LMMSE with uncalibrated LLRs, and LR with uncalibrated LLRs.

\begin{table}[htbp]\centering\small
\caption{Erroneous codewords out of 1000, IBO 3~dB.}
\label{tab:fer3}
\footnotesize
\begin{tabular*}{\tblwidth}{@{\extracolsep{\fill}}llccccccccc@{}}
\toprule
Pilots & $E_b/N_0$ & L & LMMSE/$\hat\mu$ & BA+CPE & 1-tap & LR & LR+CPE & DD-DC & L unc. & LR unc.\\
\midrule
G & 6 & 1000 & 999 & 989 & 1000 & 1000 & 1000 & 998 & 1000 & 1000 \\
G & 7 & 992 & 910 & 850 & 936 & 965 & 968 & 924 & 1000 & 995 \\
G & 8 & 936 & 607 & 458 & 672 & 770 & 768 & 603 & 994 & 972 \\
G & 9 & 720 & 276 & 176 & 351 & 473 & 470 & 268 & 962 & 889 \\
G & 10 & 425 & 99 & 55 & 135 & 179 & 169 & 97 & 918 & 719 \\
G & 11 & 202 & 27 & 10 & 46 & 70 & 64 & 17 & 773 & 500 \\
G & 12 & 98 & 9 & 2 & 23 & 25 & 18 & 7 & 641 & 328 \\
G & 13 & 31 & 5 & 1 & 6 & 8 & 7 & 2 & 486 & 222 \\
G & 14 & 15 & 0 & 0 & 3 & 4 & 2 & 0 & 415 & 167 \\
G & 15 & 3 & 1 & 0 & 1 & 1 & 0 & 0 & 293 & 116 \\
G & 16 & 3 & 2 & 1 & 2 & 2 & 1 & 1 & 250 & 77 \\
G & 17 & 4 & 0 & 0 & 1 & 1 & 0 & 0 & 242 & 105 \\
G & 18 & 2 & 0 & 0 & 0 & 1 & 0 & 0 & 184 & 64 \\
G & 20 & 0 & 0 & 0 & 0 & 0 & 0 & 0 & 161 & 48 \\
G & 24 & 0 & 0 & 0 & 0 & 0 & 0 & 0 & 143 & 49 \\
DL & 6 & 1000 & 1000 & 1000 & 1000 & 1000 & 1000 & 1000 & 1000 & 1000 \\
DL & 7 & 991 & 991 & 985 & 991 & 996 & 999 & 997 & 1000 & 1000 \\
DL & 8 & 912 & 912 & 851 & 885 & 973 & 975 & 937 & 998 & 999 \\
DL & 9 & 611 & 611 & 491 & 561 & 793 & 781 & 636 & 949 & 984 \\
DL & 10 & 264 & 264 & 160 & 216 & 423 & 415 & 273 & 852 & 908 \\
DL & 11 & 110 & 110 & 43 & 78 & 200 & 182 & 82 & 646 & 763 \\
DL & 12 & 26 & 26 & 10 & 11 & 63 & 54 & 14 & 451 & 609 \\
DL & 13 & 6 & 6 & 1 & 0 & 12 & 9 & 2 & 294 & 407 \\
DL & 14 & 0 & 0 & 0 & 0 & 6 & 3 & 0 & 183 & 295 \\
DL & 15 & 1 & 1 & 0 & 1 & 5 & 2 & 0 & 125 & 212 \\
DL & 16 & 3 & 3 & 1 & 1 & 4 & 2 & 1 & 71 & 114 \\
DL & 17 & 0 & 0 & 0 & 0 & 1 & 0 & 0 & 85 & 131 \\
DL & 18 & 0 & 0 & 0 & 0 & 1 & 0 & 0 & 57 & 78 \\
DL & 20 & 0 & 0 & 0 & 0 & 0 & 0 & 0 & 36 & 51 \\
DL & 24 & 0 & 0 & 0 & 0 & 0 & 0 & 0 & 25 & 41 \\

\bottomrule
\end{tabular*}
\end{table}

\begin{table}[htbp]\centering\small
\caption{Erroneous codewords out of 1000, IBO 6~dB.}
\label{tab:fer6}
\footnotesize
\begin{tabular*}{\tblwidth}{@{\extracolsep{\fill}}llccccccccc@{}}
\toprule
Pilots & $E_b/N_0$ & L & LMMSE/$\hat\mu$ & BA+CPE & 1-tap & LR & LR+CPE & DD-DC & L unc. & LR unc.\\
\midrule
G & 6 & 965 & 890 & 753 & 859 & 942 & 929 & 779 & 998 & 998 \\
G & 7 & 708 & 534 & 326 & 495 & 651 & 633 & 343 & 942 & 917 \\
G & 8 & 333 & 223 & 100 & 189 & 325 & 329 & 117 & 730 & 708 \\
G & 9 & 91 & 42 & 17 & 39 & 82 & 82 & 17 & 445 & 409 \\
G & 10 & 37 & 15 & 4 & 14 & 25 & 12 & 5 & 221 & 187 \\
G & 11 & 5 & 3 & 0 & 3 & 9 & 6 & 0 & 109 & 101 \\
G & 12 & 4 & 1 & 0 & 2 & 5 & 3 & 0 & 60 & 58 \\
G & 13 & 1 & 0 & 0 & 0 & 3 & 2 & 0 & 34 & 52 \\
G & 14 & 1 & 0 & 0 & 1 & 1 & 1 & 0 & 27 & 27 \\
G & 15 & 1 & 0 & 0 & 1 & 0 & 0 & 0 & 21 & 29 \\
G & 16 & 0 & 0 & 0 & 0 & 0 & 0 & 0 & 26 & 29 \\
G & 17 & 0 & 0 & 0 & 0 & 0 & 0 & 0 & 16 & 18 \\
G & 18 & 1 & 1 & 0 & 0 & 0 & 0 & 0 & 25 & 22 \\
G & 20 & 0 & 0 & 0 & 0 & 0 & 0 & 0 & 17 & 12 \\
G & 24 & 0 & 0 & 0 & 0 & 1 & 0 & 0 & 14 & 21 \\
DL & 6 & 991 & 991 & 976 & 980 & 991 & 994 & 975 & 1000 & 1000 \\
DL & 7 & 860 & 860 & 737 & 808 & 936 & 926 & 764 & 983 & 996 \\
DL & 8 & 460 & 460 & 292 & 394 & 619 & 609 & 321 & 854 & 923 \\
DL & 9 & 153 & 153 & 64 & 110 & 260 & 241 & 69 & 607 & 736 \\
DL & 10 & 43 & 43 & 15 & 23 & 72 & 58 & 11 & 323 & 408 \\
DL & 11 & 6 & 6 & 1 & 2 & 14 & 8 & 1 & 143 & 208 \\
DL & 12 & 3 & 3 & 1 & 3 & 7 & 6 & 1 & 67 & 118 \\
DL & 13 & 0 & 0 & 0 & 0 & 4 & 2 & 0 & 40 & 70 \\
DL & 14 & 0 & 0 & 1 & 0 & 1 & 1 & 0 & 18 & 35 \\
DL & 15 & 0 & 0 & 0 & 0 & 0 & 0 & 0 & 19 & 34 \\
DL & 16 & 0 & 0 & 0 & 0 & 0 & 0 & 0 & 14 & 20 \\
DL & 17 & 0 & 0 & 0 & 0 & 0 & 0 & 0 & 6 & 14 \\
DL & 18 & 1 & 1 & 0 & 0 & 0 & 0 & 0 & 11 & 19 \\
DL & 20 & 0 & 0 & 0 & 0 & 0 & 0 & 0 & 4 & 8 \\
DL & 24 & 0 & 0 & 0 & 0 & 1 & 1 & 0 & 7 & 14 \\

\bottomrule
\end{tabular*}
\end{table}

With uncalibrated LLRs, the $E_b/N_0$ for a coded BER of $10^{-3}$ at an IBO of 6~dB with impulse sounding is \codedUncLmmse~dB for LMMSE and \codedUncLR~dB for LR; the corresponding entries for the other cases are the ``uncal.'' rows of Table~\ref{tab:codedci}. No row is listed where $10^{-3}$ is not reached by 24~dB, which is the case for both receivers at an IBO of 3~dB with impulse sounding and for LR at an IBO of 3~dB with data-like pilots.

\FloatBarrier
\subsection{Independent implementation with uncalibrated LLRs}
\label{sec:indep}
An independently written coded-link simulation (true channel knowledge including phase noise, LLR scale $N_0/2$, IBO 6~dB, 3 runs $\times$ 100 codewords) gave LMMSE / LR ($N_w=1$) coded BERs of $1.65/1.47\times10^{-3}$ at 14~dB, $1.24/1.11\times10^{-3}$ at 16~dB, $1.12/0.88\times10^{-3}$ at 18~dB, $1.10/0.87\times10^{-3}$ at 20~dB, and $0.75/0.63\times10^{-3}$ at 24~dB. Both curves are flat around $10^{-3}$, so the $10^{-3}$ crossings differ by about 4~dB (LMMSE at about 21~dB and LR at about 17~dB by log-linear interpolation of these points), although the BER differs by only 10--21\%. In that implementation LR is ahead, whereas in the main study, with the same uncalibrated scaling, LMMSE reaches $10^{-3}$ earlier (\codedUncLmmse~dB against \codedUncLR~dB). The sign of the apparent coded gain therefore depends on where the error floor lies relative to $10^{-3}$ and on implementation details. Neither result is evidence about the correctors, which is why the main text uses calibrated LLRs.

\FloatBarrier
% ============================================================================
\section{Reproducibility}
\label{sec:repro}
% ============================================================================
Except for the independent check of Section~\ref{sec:indep}, all results were produced with a single NumPy implementation of the AFDM link (\texttt{engine.py}), together with Sionna 1.2.2~\cite{hoydis2022sionna} for the LDPC codec and the TDL channels. Most tables and in-text numbers of the main text and of this document are generated from the stored per-run results by one script (\texttt{gen\_all.py}; Tables~\ref{tab:fer3} and~\ref{tab:fer6} by \texttt{gen\_fer.py}). The exceptions, transcribed from archived logs or computed separately, are Tables~\Mref{tab:mu} and~\Mref{tab:complexity} of the main text, Tables~\ref{tab:params}, \ref{tab:geo}, \ref{tab:causal}, \ref{tab:esn} (except its last row), and~\ref{tab:tdlconf}, the cross-realization test (Section~\ref{sec:crossreal}), the independent-implementation values (Section~\ref{sec:indep}), and values quoted in the text from these tables (for example in Section~\Mref{sec:robust} of the main text). The scripts and per-run results (\texttt{.json}/\texttt{.npz} files) are: \texttt{exp\_unified.py} (all uncoded sweeps with common seeds; Fig.~\Mref{fig:ibo} and Table~\Mref{tab:robust} of the main text and Tables~\ref{tab:robA}--\ref{tab:robC} and~\ref{tab:ibo}), \texttt{exp\_ep.py} and \texttt{exp\_ep\_all.py} (embedded pilot), \texttt{exp\_p1.py} (Proposition~\Mref{prop:readout}), \texttt{exp\_p2d.py} and \texttt{exp\_dec2.py} (Proposition~\Mref{prop:threshold} and the variance budget), \texttt{exp\_pamis2.py} (PA-model mismatch), \texttt{exp\_p3b.py}, \texttt{sim\_win.py}, \texttt{exp\_geo.py}, \texttt{exp\_esn.py}, and \texttt{exp\_2d2.py} (Proposition~\Mref{prop:mismatch} and reservoirs), \texttt{tdl\_ci.py} and \texttt{coded\_ci.py} (confidence intervals), and archived run logs and result files for the coded link and the TDL channels (generated with Sionna on a notebook server). They are available from the corresponding author upon reasonable request.

\end{document}